\documentclass[universe,article,accept,pdftex,moreauthors]{Definitions/mdpi} 
\firstpage{1} 
\pubvolume{1}
\issuenum{1}
\articlenumber{0}
\pubyear{2024}
\copyrightyear{2024}
\datereceived{31 March 2024} 
\daterevised{13 May 2024} % Only for the journal Acoustics
\dateaccepted{17 May 2024} 
\datepublished{} 
\hreflink{https://doi.org/} % If needed use \linebreak
\usepackage{color}
\usepackage{url}
\usepackage{mathtools}
\usepackage{makecell}
\usepackage{subcaption}

\Title{Calibration Error in 21-centimeter %Attention: Title Altered.
	Global Spectrum Experiment}

\TitleCitation{Calibration Error in 21-centimeter Global Spectrum Experiments}

\Author{Shijie Sun $^{1,2}$, Eloy de Lera Acedo $^{3,4,}$*, Fengquan Wu $^{1}$,  Bin Yue $^{1}$, 
Jiacong Zhu $^{1,2}$ and Xuelei Chen $^{1,2,5,}$*}  

\AuthorNames{Shijie Sun, Eloy de Lera Acedo, Fengquan Wu,  Bin Yue, Jiacong Zhu and Xuelei Chen}

\AuthorCitation{Sun, S.; de Lera Acedo, E.; Wu, F.; Yue, B.; Zhu, J.; Chen, X.}
\address{%

$^{1}$ \quad National Astronomical Observatory, Chinese Academy of Sciences, 20A Datun Road, Beijing 100101, China; sshj@nao.cas.cn (S.S.); wufq@bao.ac.cn (F.W.); yuebin@bao.ac.cn (B.Y.); zhujiacong@nao.cas.cn (J.Z.)\\
$^{2}$  \quad School of Astronomy and Space Science, University of Chinese Academy of Sciences, Beijing 100049, China\\
$^{3}$ \quad Cavendish Astrophysics, University of Cambridge, Cambridge~CB3 0HE, UK\\
$^{4}$ \quad Kavli Institute for Cosmology in Cambridge, University of Cambridge, Cambridge~ CB3 0HE, UK\\
$^{5}$ \quad Department of Physics, College of Sciences, Northeastern University, Shenyang 110819, China
}

\corres{Correspondence: ed330@cam.ac.uk (E.d.L.A.); xuelei@cosmology.bao.ac.cn (X.C.)}

\abstract{The redshifted 21 cm line signal is a powerful probe of the cosmic dawn and the epoch of reionization. The global spectrum can potentially be detected with a single antenna and spectrometer. However, this measurement requires an extremely accurate calibration of the instrument to facilitate the separation of the 21 cm signal from the much brighter foregrounds and possible variations in the instrument response. Understanding how the measurement errors propagate in a realistic instrument system and affect system calibration is the focus of this work. We simulate a 21 cm global spectrum observation based on the noise wave calibration scheme. We focus on how measurement errors in reflection coefficients affect the noise temperature and how typical errors impact the recovery of the 21 cm signal, especially in the frequency domain.  Results show that for our example set up, a typical vector network analyzer (VNA) measurement error in the magnitude of the reflection coefficients of the antenna, receiver, and open cable, which are 0.001, 0.001, and 0.002  (linear), respectively, would result in a 200 mK deviation on the detected signal, and a typical measurement error of $0.48^\circ$, $0.78^\circ$, or $0.15^\circ$ in the respective phases would cause a 40 mK deviation. The VNA measurement error can greatly affect the result of a 21 cm global spectrum experiment using this calibration technique, and such a feature could be mistaken for or be combined with the 21 cm signal.}


\begin{document}

%%%%%%%%%%%%%%%%%%%%%%%%%%%%%%%%%%%%%%%%%%
% \setcounter{section}{-1} %% Remove this when starting to work on the template.
% \section{How to Use this Template}

% The template details the sections that can be used in a manuscript. Note that the order and names of article sections may differ from the requirements of the journal (e.g., the positioning of the Materials and Methods section). Please check the instructions on the authors' page of the journal to verify the correct order and names. For any questions, please contact the editorial office of the journal or support@mdpi.com. For LaTeX-related questions please contact latex@mdpi.com.%\endnote{This is an endnote.} % To use endnotes, please un-comment \printendnotes below (before References). Only journal Laws uses \footnote.

% % The order of the section titles is: Introduction, Materials and Methods, Results, Discussion, Conclusions for these journals: aerospace,algorithms,antibodies,antioxidants,atmosphere,axioms,biomedicines,carbon,crystals,designs,diagnostics,environments,fermentation,fluids,forests,fractalfract,informatics,information,inventions,jfmk,jrfm,lubricants,neonatalscreening,neuroglia,particles,pharmaceutics,polymers,processes,technologies,viruses,vision

\section{Introduction}
\label{sec:introduction}

\textls[-5]{The redshifted 21 cm line of neutral hydrogen, potentially observable at low radio 
frequencies (50--200 MHz), should be a powerful probe of the physical conditions of the inter-galactic medium during the cosmic dawn (CD) and the epoch of 
reionization (EoR)~\citep{madau199721,Chen2004,Chen2008,furlanetto2006cosmology,barkana2016rise}.}

The global (all-sky averaged) spectrum of the redshifted 21 cm brightness temperature can be measured with a single antenna and wide band spectrometer. 
It is a direct approach with a high precision instrument. 
The experimental efforts include the Experiment to Detect the Global EoR Signature (EDGES,~\cite{bowman2008toward,2010}), the Broadband Instrument for Global HydrOgen ReioNization Signal (BIGHORNS,~\cite{sokolowski2015bighorns}), the Shaped Antenna measurement of the background RAdio Spectrum (SARAS,~\cite{patra2013saras,saras2018,saras3}), the Probing Radio Intensity at high-Z from Marion (${\rm PRI^ZM}$,~\cite{philip2019probing}), Radio Experiment for the Analysis of Cosmic Hydrogen (REACH,~\cite{de2019reach,cumner2021radio}), the Large-aperture Experiment to Detect the Dark Age (LEDA,~\cite{price2018design}), Sonda Cosmologica de las Islas para la Deteccion de Hidrogeno Neutro (SCI-HI,~\cite{2014}), Cosmic Twilight Polarimeter (CTP,~\cite{2019}), and Mapper of the IGM Spin Temperature (MIST,~\cite{MIST-web}), which use interferometric measurements to help estimate instrumental and foreground parameters. There are also experiments using short spacing interferometers, such as the Short spacing Interferometer Telescope probing cosmic dAwn and epoch of ReionizAtion (SITARA)~\cite{Thekkeppattu_2022}.
Considering the measurement difficulty from most ground sites at low frequencies due to ionosphere refraction and reflection of broadband radio frequency interference (RFI), there are some proposals to go to the far side of the moon, for example the Dark Ages Polarimetry PathfindER in low lunar orbit (DAPPER,~\cite{burns2021global}), Dark Ages Radio Explorer project (DARE,~\cite{burns2017space}), and the Discovering the Sky at Longest wavelength project (DSL,~\cite{Chen:2020lok}). 

\textls[-25]{The EDGES experiment has reported the detection of a 500 mK deep absorption feature centered at 78 MHz \citep{bowman2018absorption}, which may be associated with the signature of cosmic dawn. This absorption strength is much stronger than the prediction of the standard model,  so its cosmological interpretation may require new physics or astrophysics mechanisms, e.g., cooling mechanism with exotic dark matter particles\cite{Barkana2018Natur}, or extra radio background~\cite{21cm_moreCRB_1802.07432,BH_CRB_1803.01815}. However, it has also been questioned whether this feature is true or arises from some systematical errors~\cite{hills2018, Bradley_2019}, and 
the measurement taken by the SARAS-3 experiment has not detected such an absorption feature \citep{saras2021}. }

Observing the global 21 cm signal is very challenging, as the low-frequency radio sky is dominated by intense synchrotron emission from our own galaxy, which is more than four orders of magnitude brighter than the signal. There are also galactic free--free emissions, numerous compact radio sources such as supernova remnants, quasars, and radio galaxies, which contribute to the total received power. Extremely high sensitivity and large dynamic ranges are required to discern the small 21 cm signature in the spectrum. Moreover, the cosmic dawn 21 cm signal has an unknown but generally broad shape. Its detection requires a good understanding of the instrument response, which must be determined by a calibration procedure. Any frequency-dependence in instrumental gain, noise spectrum, antenna beam shape, and ionospheric effect, if not properly accounted for,  can affect the measurement result. In particular, the reflections of electric signal at the interfaces of the various components of the measurement system are frequency-dependent and could not be neglected. 

To derive an accurate global spectrum, a calibration with very high precision is required. The calibration of an instrument can be based on either the measurement of an external standard source, or by the internal calibration mechanisms designed into the system. For the global spectrum experiment, calibration using an external standard source is difficult, because its antenna has a very wide beam, covering essentially the whole sky above the horizon. Although some crude calibration could be achieved with a model of sky radiation and helped by the variation in sky due to Earth rotation, as, for example, used by the SCI-HI experiment~\cite{2014}, the precision of such calibration is limited by the large uncertainties in the sky model and the beam model. It is also conceivable that one could try to use an artificial external source, but again, this is complicated by the antenna response, which is not isotropic. To mimic the response to the whole sky, the whole apparatus can, in principle, be enclosed inside an anechoic chamber kept at a constant temperature. Ideally, the instrument is thus well-isolated and immersed in the black body radiation of the chamber, which can serve as the standard source for calibration \endnote{Private communication from Prof. Jeffery Peterson, 2012.}. However, it is very challenging to build such an anechoic chamber with sufficient shielding and absorbing power at low frequency and keeping all its parts at the same temperature. 

Presently, a more practical approach to high-precision calibration is based on internal calibrators and the  noise wave formulation of the electric circuit system \citep{1129303}. This approach was pioneered by the EDGES group \citep{rogers2012absolute} and subsequently adopted by most experiments.  In this formulation,  the linear radio frequency (RF) devices are characterized by a few fundamental wave parameters, which can be determined experimentally by measuring the reflection coefficients using the vector network analyzer (VNA). It is then possible to relate the noise power measured by the instrument to the sky temperature and give an estimate of the error in the measured results. 

To assess the uncertainties of the measurement, in addition to the thermal noise, one also needs to consider the possible systematic errors that may bias the result. This is the main aim of the present work. Previously, \citet{Monsalve_2017} and \citet{monsalve2017results} presented the calibration strategy of the EDGES experiment and analyzed the propagation of uncertainties in the receiver calibration. In a recent work \citep{2021MNRAS.tmp.1408R}, a Bayesian approach to the noise wave formulation problem (the one used by the REACH experiment \citep{de2019reach}) is presented. In this work, we investigate this problem with realistic models and a detailed characterization of the systematic errors. In particular, we consider the VNA measurement error, which depends on frequency, and adopt a comprehensive model for VNA measurement uncertainties, including all measurement errors in the VNA measurement procedure and in calibration standards, to estimate their impacts on the spectrum measurement system. Furthermore, we also study how the error affects the extracted 21 cm signal, i.e., the non-smooth component of the sky spectrum. In the simulation, we have adopted the parameter values we found in our actual experiment work. While there are many possible instrument designs, and many hardware components with different characteristics, the differences are not that large. We have used typical values, although there is still room for refinement.

The rest of this paper is organized as follows: 
Section~\ref{sec:models and simulation methods} describes the foreground model and the 21 cm signal model, introduces the noise wave formulation (some derivation given in Appendix~\ref{appa}) and the calibration scheme based on it, and establishes a framework for the cosmological spectrum measurement system simulation, with an emphasis on measurement uncertainties from the vector network analyzer (VNA) (details given in Appendix~\ref{appb}) and its impact on cosmological spectrum measurement. Section~\ref{sec:Error propagation} presents the results of a simulation of the global spectrum measurement system, adding VNA measurement uncertainties and other generic systematic errors. How different errors impact the signal recovery are then discussed. Section~\ref{sec:conclusions and future work} summarizes our conclusions and discusses future~work.

\section{Models and Simulation Methods}
\label{sec:models and simulation methods}

\subsection{Foreground and 21 cm Spectrum Model}

The galactic foreground radiation comes from a variety of complex physical processes~\citep{de2008model}. At low frequency, the synchrotron radiation from cosmic ray electrons spiraling in the galactic magnetic field dominates. In this paper, following \citet{bowman2018absorption}, the foreground received by the ground-based experiment is modeled using polynomials of frequency,with five terms based on the known spectral properties of the galactic synchrotron spectrum and Earth’s ionosphere \citep{2015Radiometric}, as:  
\begin{equation}
\begin{aligned}
    T_F(\nu) \approx & a_0\left(\frac{\nu}{\nu_c}\right )^{-2.5} + a_1\left(\frac{\nu}{\nu_c}\right )^{-2.5} \log_{}({\frac{\nu}{\nu_c}}) +\\ 
    & a_2\left(\frac{\nu}{\nu_c}\right )^{-2.5} \left [ \log_{}({\frac{\nu}{\nu_c}}) \right ] ^2+a_3\left(\frac{\nu}{\nu_c}\right )^{-4.5} 
    +a_4\left(\frac{\nu}{\nu_c}\right )^{-2}
    \label{eq:foreground}
\end{aligned}
\end{equation}
where $T_F(\nu)$ is the brightness temperature of the foreground emission,  $\nu$ is the frequency, 
$\nu_c$ is the center frequency of the observed band, and the coefficients $a_n$ are fitted to the data. 

\begin{figure}[H]
%\centering
	\includegraphics[width=0.9\columnwidth]{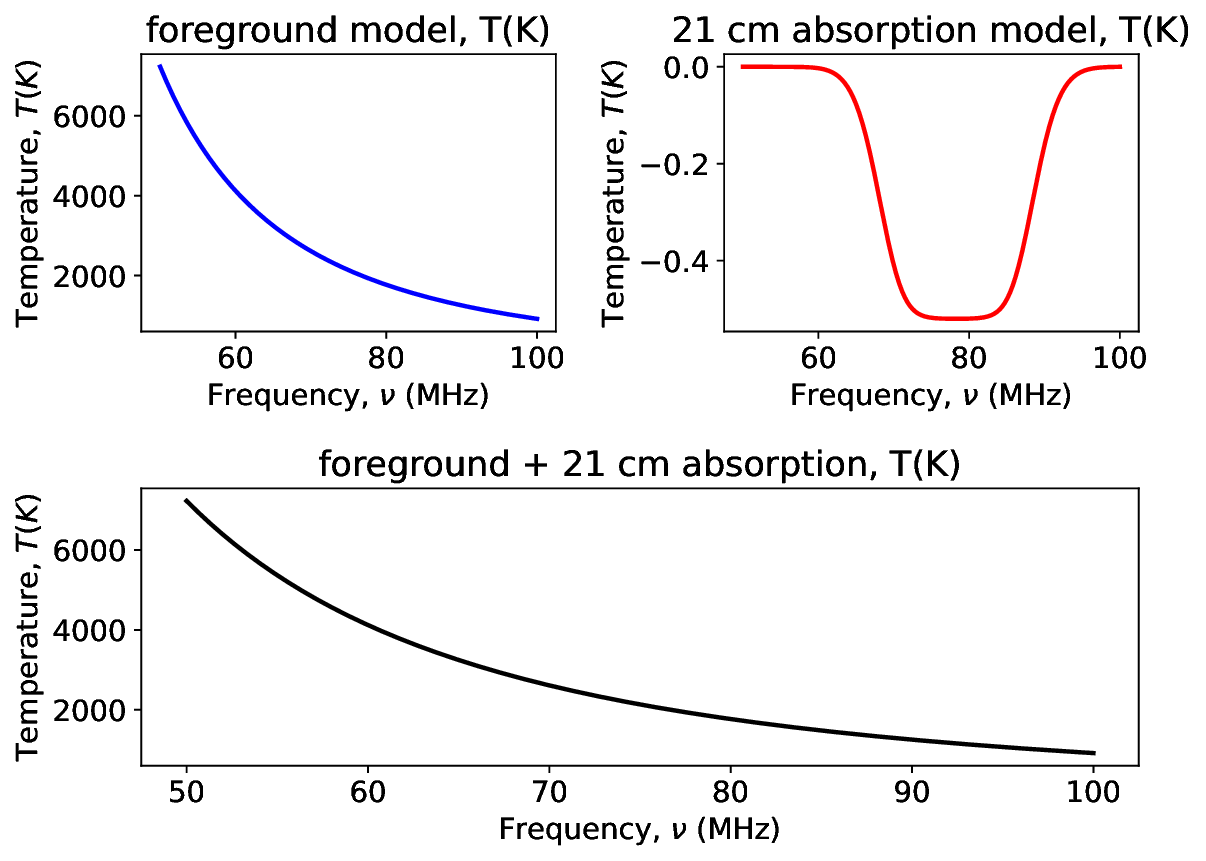}
    \caption{\textbf{Top left:} Foreground model. Galactic synchrotron emission dominates this band, yielding a smooth power-law-like spectrum that decreases from approximately 7000 K at 50 MHz to approximately 1000 K at 100 MHz. \textbf{Top right:} 21 cm absorption model, a negative flattened Gaussian centered at 78 MHz, with bandwidth of approximately 30 MHz. \textbf{Bottom}: Combined spectrum of foreground and 21 cm absorption. The foreground is approximately five orders higher than the 21 cm absorption feature.} 
    \label{fig1}
\end{figure}

The brightness temperature of the redshifted 21 cm signal from the early universe is frequency dependent \citep{madau199721}. The predicted spectral signature is a relatively broadband absorption  (during dark ages and cosmic dawn) and emission (during EoR) signal from 50--200 MHz ($30 \le z \le 6$), with a peak absolute absorption amplitude between 10 and 250 mK, which is dependent on the particular model of the first stars/galaxies \citep{cohen2017charting}.
In this paper, the 21 cm absorption profile during the cosmic dawn is modeled as a negative flattened Gaussian format \citep{bowman2018absorption}:
\begin{equation}
    T_{21}(\upsilon)=-A\left(\frac{1-e^{-\tau^B}}{1-e^{-\tau}}\right)
	\label{eq:21 cm absorption}
\end{equation}
where $A$ is the absorption amplitude, and 
\begin{equation}
    B=\frac{4(\upsilon-\upsilon_0)^2}{w^2}\log\left[-\frac{1}{\tau}\log\left(\frac{1+e^{-\tau}}{2}\right)\right]
    \label{eq:21 cm absorption 2}
\end{equation}
$\upsilon_0$ is the center frequency, $w$ is the full-width at half-maximum (FWHM), and $\tau$ is a flattening factor.
This is a mathematical description of the line profile, not a description of the physics that create the 21 cm absorption profile. We adopt the EDGES measured 21 cm absorption spectrum as our 21 cm spectrum fiducial model. However, we argue that the results of this paper, i.e., the impact of systematic error, are not sensitive to a particular 21 cm signal model or foreground model. The input value of the parameters of the foreground and 21 cm absorption signal model used in this work are listed in Table.~\ref{tab1}. The foreground, 21 cm absorption, and the combined foreground and 21 cm absorption signal are shown as 
Figure~\ref{fig1}.

\begin{table}[] 
\caption{The input value of the foreground and 21 cm signal model parameters.}
\label{tab1}
\newcolumntype{C}{>{\centering\arraybackslash}X}
\begin{tabularx}{\textwidth}{CCCCC}
\toprule
%\centering
{$a_0$}&{$a_1$}&{$a_2$}&{$a_3$}&{$a_4$} \\
\midrule
{1284}&{570}&{$-$1240}&{753}&{98} \\
\midrule
{$A$}&{$\upsilon_0$}&{$w$}&{$\tau$} \\
\midrule
{0.52}&{78.3}&{20.7}&{6.5} \\
\bottomrule
\end{tabularx}
\end{table}

\subsection{The Noise Wave Formulation}

Unlike the artificially generated signals, which often have regular wave forms, the naturally produced astronomical signals often appear to be varying randomly and can be well-characterized by their power or temperature. In global spectrum measurement instruments, where the radio waves are converted to electrical signals, the astronomical signals are joined by the noise generated within the instruments. At the interfaces of different components of the electronic system, part of the signal propagates down to the next component, and part of the signal is reflected back if the impedance of the two parts is not perfectly matched. Moreover, the receiver is an active device, it does not only receive the signal passively at its front end but also emits noises. 

The noise wave concept \citep{1129303} provides a description of the propagation of such noise-like signals in the electronic system. This formulation describes the noise behavior of linear RF devices with a few measurable parameters of the system, such as the reflection coefficients and physical temperatures of the calibrators. 
In this formulation, we quantify the power of the electrical signal by a noise temperature. As the signal propagates through the system and gets reflected back and forth at the junctions of the various components of the system due to impedance mismatches, the noise power and temperature are accordingly transferred. The signal power at the receiver, as induced by the signal from the antenna, can be written as  
$P= k T_{\rm ant} \Delta \nu$, where $k$ is the Boltzmann constant, $\Delta\nu$ is the bandwidth, and $T_{\rm ant}$ is the so-called antenna temperature, which, according to the noise wave model, is related to the averaged sky temperature $T_{\rm sky}$ by
\begin{equation}
    \begin{aligned}
        T_{\rm ant}&=T_{\rm sky}(1-|\Gamma_{\rm ant}|^2)|F_{\rm ant}|^2+T_u|\Gamma_{\rm ant}|^2|F_{\rm ant}|^2+\\
        &(T_c \cos(\phi)+T_s \sin(\phi))|\Gamma_{\rm ant}||F_{\rm ant}|+T_0,
        \label{eq:noise wave}
    \end{aligned}
\end{equation}
where the transfer function of the noise wave reflected between the antenna and receiver is
\begin{equation}
    F_{\rm ant}=\frac{ (1-|\Gamma_{\rm rec}|^2)^\frac{1}{2} }{ 1-\Gamma_{\rm ant}\Gamma_{\rm rec} },
    \label{eq:F term}
\end{equation}
and the reflection coefficients of the antenna and receiver are, respectively,
\begin{equation}
    \Gamma_{\rm ant}=\frac{Z_{\rm ant}-Z_0 }{Z_{\rm ant}+Z_0},
    \label{eq:antenna temperature gamma antenna}
\end{equation}
\begin{equation}
    \Gamma_{\rm rec}=\frac{Z_{\rm rec}-Z_0}{Z_{\rm rec}+Z_0},
    \label{eq:antenna temperature gamma receiver}
\end{equation}
here, $Z_0$ denotes the characteristic impedance of the transmission line and can be viewed as a reference impedance, and $T_0$ represents the receiver noise offset. The noise reflected from the antenna reenters the receiver with phase $\phi$, which is equal to the phase of $\Gamma_{\rm ant}F_{\rm ant}$. $T_u$ is the uncorrelated portion of the receiver noise reflected back from the antenna, and $T_c$ and $T_s$ are the cosine and sine components of the correlated receiver noise reflected back by the antenna. The derivation of this formula is given in Appendix \ref{appa}.

The reflection coefficient in this formula depends on the impedance of the source and the network connected to the source. For a typical two port network with finite impedance, $|\Gamma_{s}|=1$ for an open circuit source ($Z=\infty$) or a shorted source ($Z=0 \Omega$),  while $|\Gamma_{s}|=0$ for a matched load with the same impedance as the network. $T_u$, $T_c$, $T_s$, and $\phi$ can be determined from measurements.

%%%%%%%%%%%%%%%%%%%% next section %%%%%%%%%%%%%%%%%%%%%

\subsection{The Global Spectrum Measurement System}

\begin{figure}[H] 
	\includegraphics[width=\linewidth]{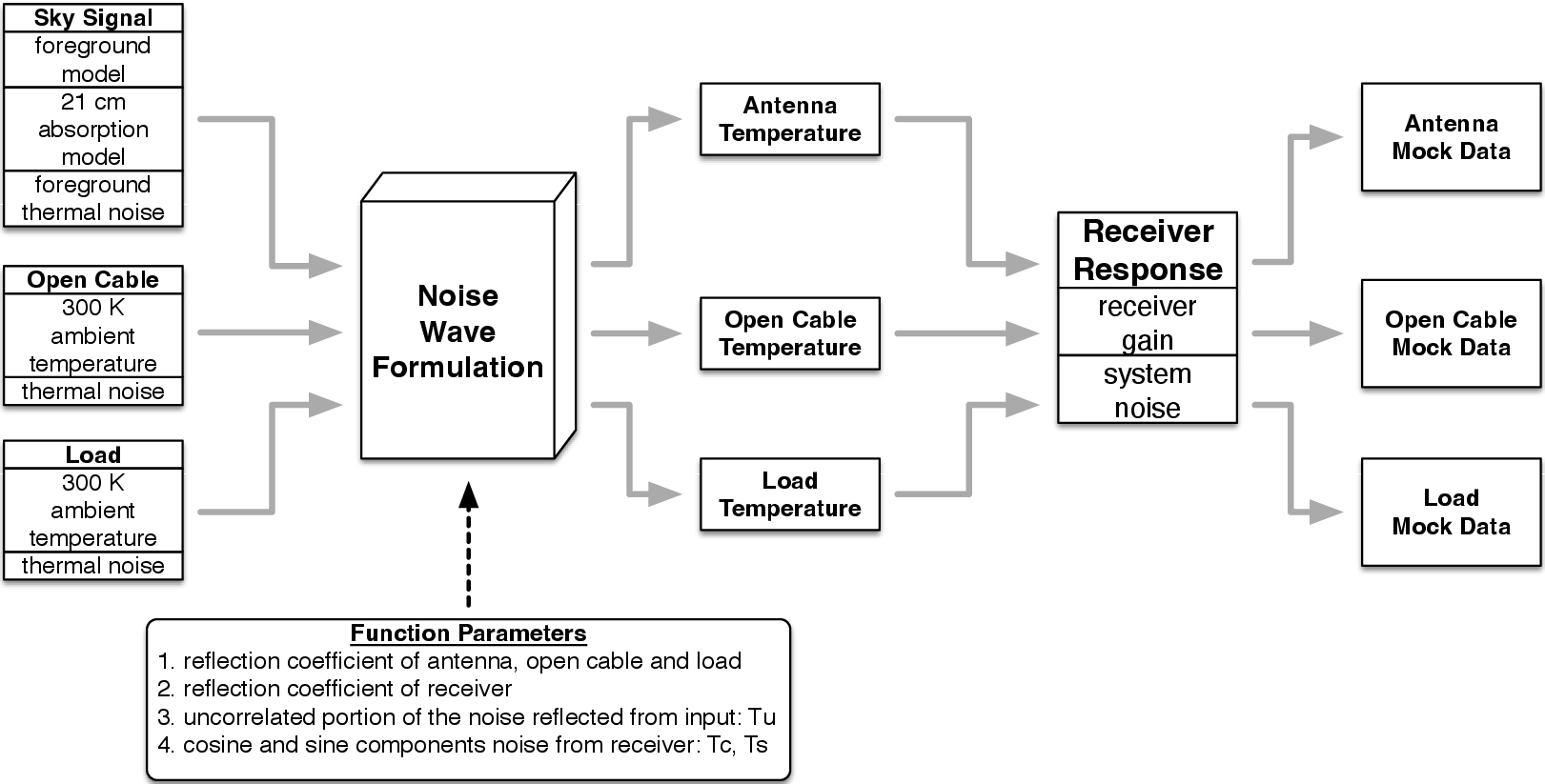}
    \caption{Simulation block diagram for global spectrum measurement and calibration system. }
    \label{fig2}
\end{figure}

The block diagram of the global spectrum measurement instrument analyzed in this paper is shown in Figure~\ref{fig2}. This system can be switched to one of three inputs: an antenna for observation of the sky signal, an open cable, and a 50 $\Omega$ resistance load, the latter two inputs are used for system calibration. The receiver is modeled as a two-port network system. We use a method similar to the one used by the EDGES experiment to perform calibration and signal reconstruction. Here, we use a 50 $\Omega$ ambient resistor load to calibrate the system response and assume a stable and precise 300 K ambient temperature load for absolute temperature calibration. We note that both an ambient and a hot calibrator are used in the EDGES experiment, which can provide calibration for two parameters---the frequency-dependent temperature scale factor and the offset. In the present work, we focus on the impact of the VNA measurement error; therefore, here, we only use an ambient load for temperature reference, which simplifies the description and does not significantly affect our estimates. In actual experiments, however, it may be necessary to introduce both the ambient and hot~calibrators.

An open cable is used to solve the noise wave parameters in this simulation. Here, we adopted a simple model of the system, with $T_u, T_c,$ and $T_s$ as linear functions of frequency. Results show that the fitting residue is less than 3 mK when using only an open cable to solve the noise wave parameters. In a real experiment, the system could be more complicated, e.g., the system temperature could have a more complicated dependence on the frequency. An additional shorted cable can provide an independent calibration measurement and, thus, help improve the accuracy of the noise wave parameter determination.

We focus on how the systematic errors propagate in the global spectrum measurement system, as well as their impact on the recovered signal. We first consider the time-independent case; the time variation in the system parameters such as the gain, the noise excess temperature, and the ambient temperature are discussed in the latter sections.

When an antenna is connected to the receiver, the sky signal is fed into the system, but if the antenna output impedance and the receiver input impedance are not a perfect match, a fraction of the signal power would be reflected back to the antenna. Moreover, the receiver also produces noise that propagates back to the antenna. The receiver noise can be separated into a component that is correlated with the receiver output and a component that is uncorrelated \citep{rogers2012absolute}. The uncorrelated noise power depends only on the magnitude of the antenna reflection, while the correlated noise power depends also on the phase. The antenna temperature is related to the sky noise power, as given by Equation~(\ref{eq:noise wave}). 

For receiver calibration, the receiver input can also be switched to a 50 $ \Omega$ ambient resistor load or an open cable. With the 50 $\Omega$ resistor load as input, the sky temperature $T_{\rm sky}$ is replaced by the ambient temperature $T_{\rm amb}$, and the receiver noise temperature can be written~as

\begin{equation}
\begin{aligned}
    T_{\rm load}&=T_{\rm amb}(1-|\Gamma_{\rm load}|^2)|F_{\rm load}|^2+T_u|\Gamma_{\rm load}|^2|F_{\rm load}|^2+\\
    &(T_c \cos(\phi)+T_s \sin(\phi))|\Gamma_{\rm load}||F_{\rm load}|+T_0, 
    \label{eq:load temperature}
\end{aligned}
\end{equation}
where 
\begin{equation}
    F_{\rm load}=\frac{ (1-|\Gamma_{\rm rec}|^2)^\frac{1}{2} }{ 1-\Gamma_{\rm load}\Gamma_{\rm rec} }, 
    \label{eq:load temperature F}
\end{equation}
\begin{equation}
    \Gamma_{\rm load}=\frac{Z_{\rm load}-Z_0}{Z_{\rm load}+Z_0}, 
\end{equation}
$\Gamma_{\rm rec}$ is given by Equation~(\ref{eq:antenna temperature gamma receiver}) since the same receiver is used, $\phi$ is the phase of $\Gamma_{\rm load}F_{\rm load}$, and $T_{\rm amb}$ is the ambient temperature of the resistance load. For a realistic experiment system, $Z_{\rm load}\approx 50\ \Omega$, then $\Gamma_{\rm load}\approx 0$. Equation~(\ref{eq:load temperature}) is reduced to
\begin{equation}
\begin{aligned}
    T_{\rm load}=T_{\rm amb}(1-|\Gamma_{\rm rec}|^2)+T_0
    \label{eq:load temperature approx}
\end{aligned}
\end{equation}

Similarly, when connected to the open cable, the receiver noise temperature is given~by
\begin{equation}
\begin{aligned}
    T_{\rm open}&=T_{\rm amb}(1-|\Gamma_{\rm open}|^2)|F_{\rm open}|^2+T_u|\Gamma_{\rm open}|^2|F_{\rm open}|^2+\\
    &(T_c \cos(\phi)+T_s \sin(\phi))|\Gamma_{\rm open}||F_{\rm open}|+T_0, 
    \label{eq:open temperature}
\end{aligned}
\end{equation}
where 
\begin{equation}
    F_{\rm open}=\frac{ (1-|\Gamma_{\rm rec}|^2)^\frac{1}{2} }{ 1-\Gamma_{\rm open}\Gamma_{\rm rec} }, 
    \label{eq:open temperature F}
\end{equation}
\begin{equation}
    \Gamma_{\rm open}=L_c\frac{Z_{\rm in}-1}{Z_{\rm in}+1}, 
    \label{eq:gamma open}
\end{equation}
where $L_c$ is the cable loss factor $L_c=10^{-\frac{l_c}{10}}$ , $l_c$ is the one-way cable loss, which is a function of 
frequency $\nu$. For open cable, $Z_{\rm in}=-jZ_0\cot\beta l$.
The cable acts like an antenna looking at an isotropic sky with temperature equal to the physical temperature of the cable.
The noise wave parameters $T_u$, $T_c$, and $T_s$ can be solved from these calibration measurements.

\begin{figure}[H]
%\centering
\includegraphics[width=1\linewidth]{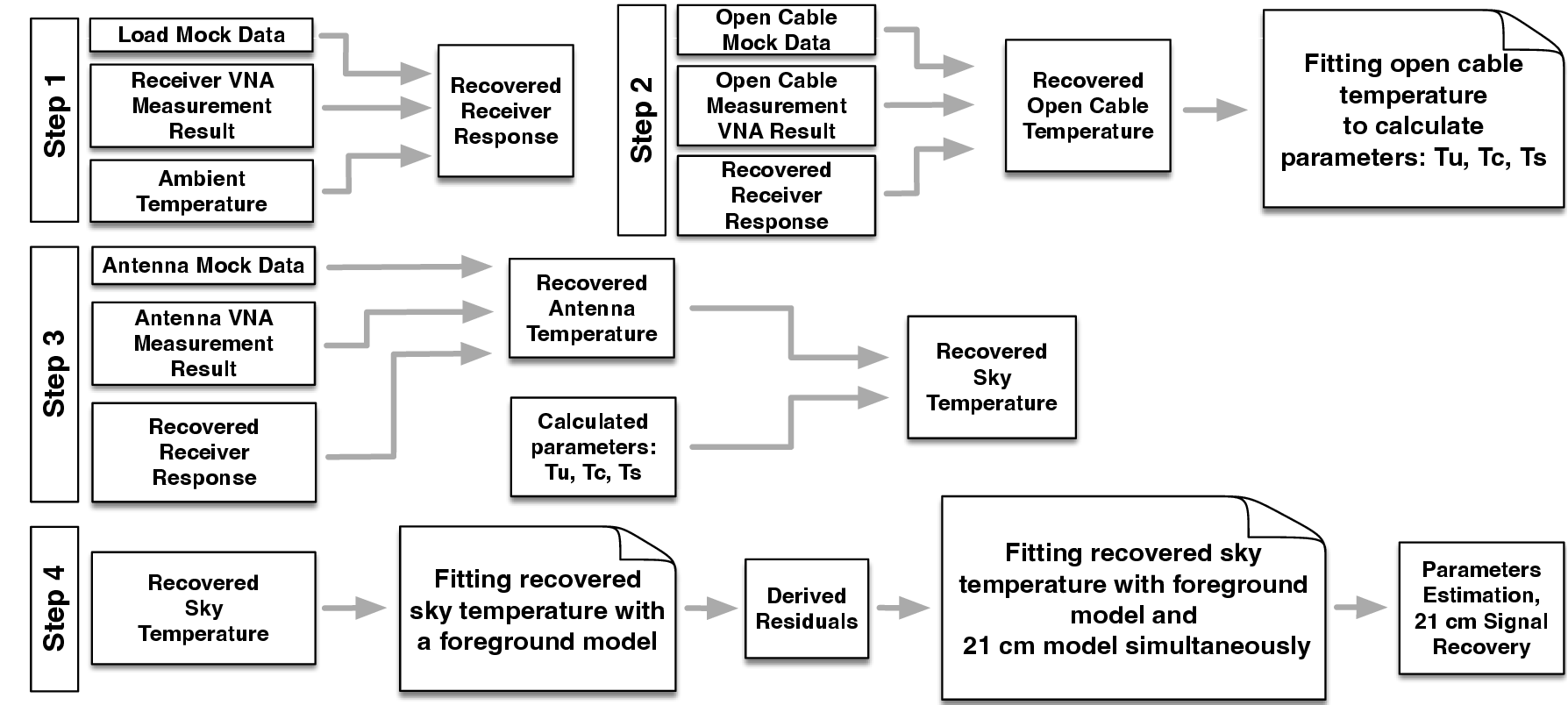}
\caption{Simulation block diagram of the sky signal reconstruction for global spectrum measurement~system. }
\label{fig3}
\end{figure}

In Figure~\ref{fig3}, we plot the block diagram of the sky signal reconstruction procedure for this global spectrum measurement system. The 21 cm spectrum signal is recovered from the observation and calibration mock data in 4 steps: (1) Calibrate for the receiver response using the load calibration (mock) data, the receiver reflection coefficient measurement data, with assumed ambient temperature for the load, and receiver noise temperature. (2) Recover the open cable temperature $T_{\rm open}$, fitting it to the noise wave formulation with the reflection characteristics of the open cable and receiver, then solve for noise wave parameters $T_u, T_c,$ and $T_s$. (3) Using the system parameters obtained in the first two steps to recover antenna temperature from the raw data, reconstruct the sky temperature. (4) Fit the sky temperature with foreground model only and with both foreground and 21 cm model simultaneously, then derive the foreground and 21 cm model parameters.

%%%%%%%%%%%%%%%%%%%%%%%%%%%%%%%%%%%%%%%%%%%%%%%%%%%%%%%%%%%%%%%%%%%%%
\subsection{Hardware Models}

We assume the antenna to be a wideband blade dipole similar to the one used by the EDGES experiment, though the exact parameters may differ. The antenna response is obtained by an electromagnetic field simulation using the CST STUDIO SUITE (\cite{CST})~software. 

\begin{figure}[H] 
%\centering
	\includegraphics[width=0.8\columnwidth]{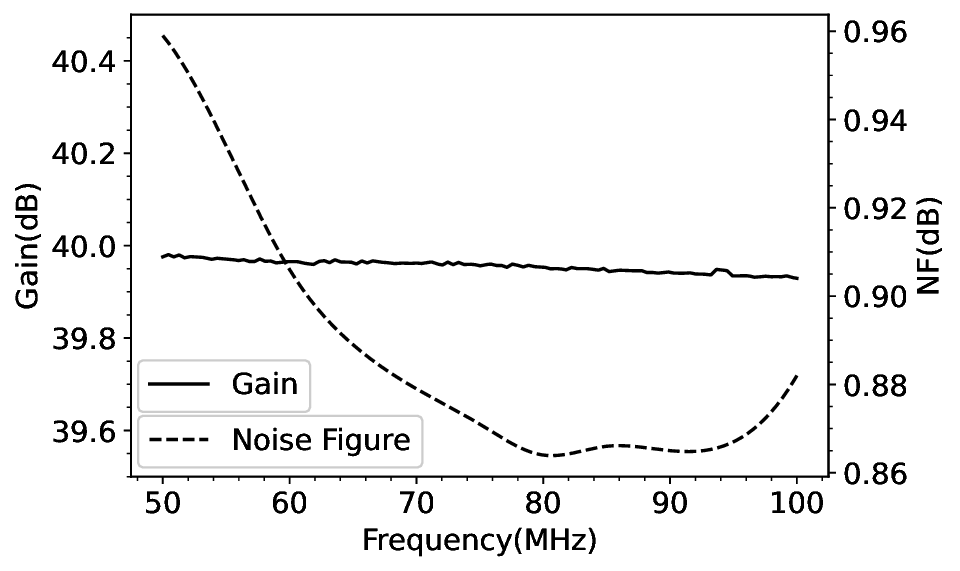}
    \caption{The Gain and Noise Figure of the LNA receiver. }
    \label{fig4}
\end{figure}

We use some typical and representative models and parameters for receiver and system settings to be as close as possible with the actual observation systems, though it is understood that for different brands and types, the parameter value could vary. To obtain concrete estimates of the noise, we assume the following design and components. The custom-made low noise amplifier (LNA) receiver covers the 30--250 MHz band and is optimized for the low noise figure and flatness of $S_{11}$ and $S_{22}$. The Wantcominc WHM0003AE~\citep{wantcominc} transistor is used for the input stage as well as the second amplification stage. This is a wideband, high linearity SMT packaged amplifier with exceptional gain flatness design. The gain and noise figure of the LNA are plotted in Figure~\ref{fig4} as a function of frequency. The noise figure is typically 0.70 dB. Measurements of the LNA show an average $S_{11}$ of $-$21.6 dB, with a flatness better than 0.4 dB within the band, and the average $S_{22}$ is$ -$20.0 dB, with a flatness better than 0.1 dB across 50--100 MHz. The LNA typically offers a 40 dB gain with a flatness of 0.05 dB, and the noise figure is less than 0.95 dB across the band 50--100 MHz. 

An open-terminated cable of 5 m length is used for internal calibration of the system. The open cable model is based on a precision test cable assembly, its impedance is 49.6 $\Omega$, and the velocity of propagation in the cable is 83\% of the speed of light; a first-order polynomial of frequency is used as the attenuation coefficient, i.e., 0.24 dB/m @ 50 MHz, 0.30 dB/m @ 100 MHz. The load used for receiver calibration is modeled as a pure impedance terminator; its impedance is set as 50 $\Omega$.

The reflection coefficients ($S_{11}$) of the antenna, the LNA receiver, and the open cable are shown in {Figure}~\ref{fig6} as a function of frequency. These are complex quantities so both the magnitudes and phases are shown, and the expected measurement errors  (to be discussed in the next section) are also plotted.

\subsection{VNA Measurement Error Model}
\label{sec:VNA measurement error model}

A vector network analyzer (VNA) can be used to measure the $S$ parameters of a one-port or two-port system. In a 21 cm global spectrum measurement, a VNA is employed to measure  the reflection coefficients \citep{rogers2012absolute}. The measurement accuracy of the reflection coefficients ($S_{11}$) of the antenna, cable, and receiver are very important for our purpose. As a measurement instrument, VNA has
errors that are caused by imperfections within the network analyzer; these errors are referred to as measurement errors. Measurement errors can be classified into three groups: systematic, random, and drift and stability. A more detailed description of the source of a VNA error is given in Appendix~\ref{appb}.

For a given set of  VNA, standard calibrator, and cable, the measurement error or uncertainty in the reflection coefficient ($S_{11}$) can be estimated using the  VNA error model, as shown in Figure~\ref{fig:vna signal flow graph}, with the magnitude/phase given in Equations~(\ref{eq:error magnitude}) and ~(\ref{eq:error phase}). Below, we take the Keysight$^{\rm TM}$ N5247B PNA-X series vector network analyzer and the corresponding 85050C precision mechanical calibration kit as the components used in our model. Table~\ref{tab2} lists the typical specifications for directivity, source match, load match, reflection tracking, and dynamic accuracy of this type of VNA. These specifications are for typical measurement setups: 10 Hz IF bandwidth, isolation calibration with an averaging factor of 8, and the specifications of the systematic effects are guaranteed values, and they should be the maximum values in the frequency domain. These are estimated with 
the {\tt vector network analyzer uncertainty calculator} script supplied by its maker \citep{VNAuncertainty}.  

\begin{table}[H] 
  \caption{Corrected system performance of the Keysight$^{\rm TM}$ N5247B PNA-X series vector network analyzer with the 85050C standard calibration kit (50--500 MHz). Dynamic accuracy is specified by the condition of $-$10 dBm input power at 50 MHz.}
\newcolumntype{C}{>{\centering\arraybackslash}X}
\begin{tabularx}{\textwidth}{CCCC}
\toprule
      Directivity&Source Match&Load Match \\ 
      \midrule
      Mag & Mag & Mag \\
      \midrule
      48 dB&40 dB&47 dB \\
       \midrule
      \multicolumn{2}{@{}c@{}}{Reflection Tracking} & \multicolumn{2}{@{}c@{}}{Dynamic Accuracy} \\
     \midrule
      Mag & Phase & Mag & Phase \\
      \midrule
      $\pm 0.0030$ dB&$\pm 0.020^\circ$&0.009 dB&$0.08^\circ$ \\
      \bottomrule
    \end{tabularx}\label{tab2}
\end{table}

\begin{figure}[H]  
%\centering
\includegraphics[width=\columnwidth]{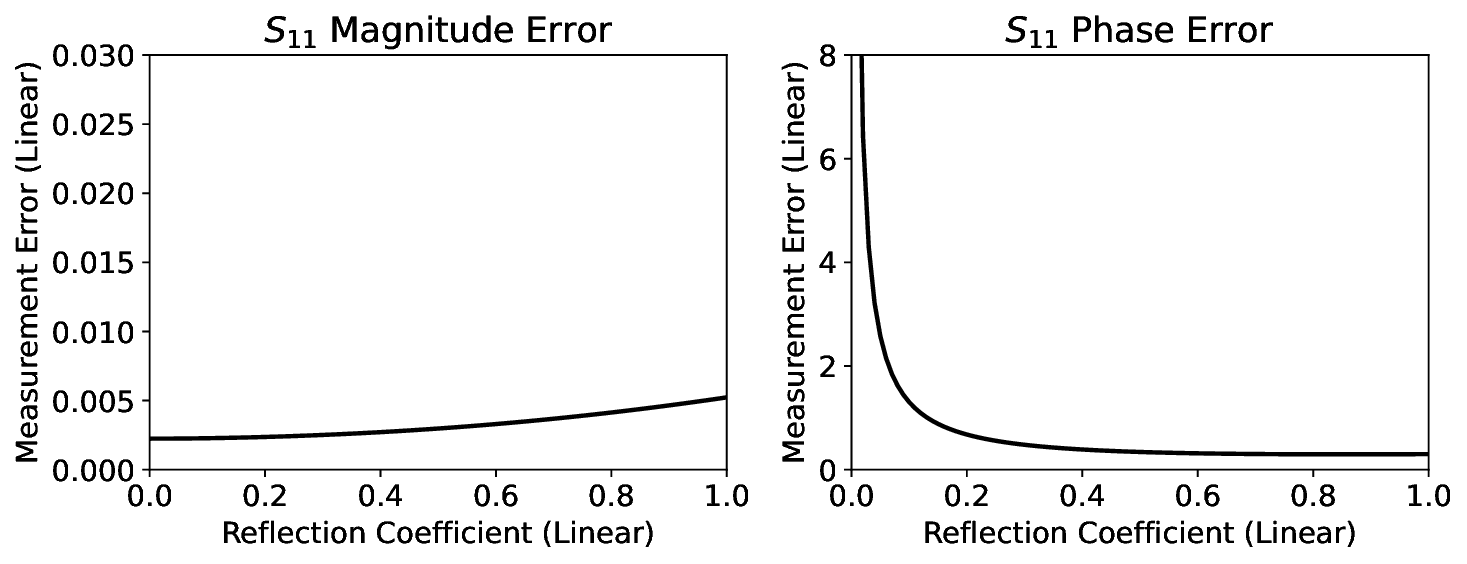}
    \caption{The VNA Reflection Coefficients Measurement Errors  (Left: magnitude, Right: phase) for a Keysight N5247B PNA-X Microwave Network Analyzer, with IF bandwidth 1Hz. Average factor 2000, calibration power, and measurement are all set to 0 dBm. The 85050C precision mechanical calibration kit is used for VNA calibration.}
    \label{fig5}
\end{figure}

\begin{figure}[H] 
\includegraphics[width=0.7\columnwidth]{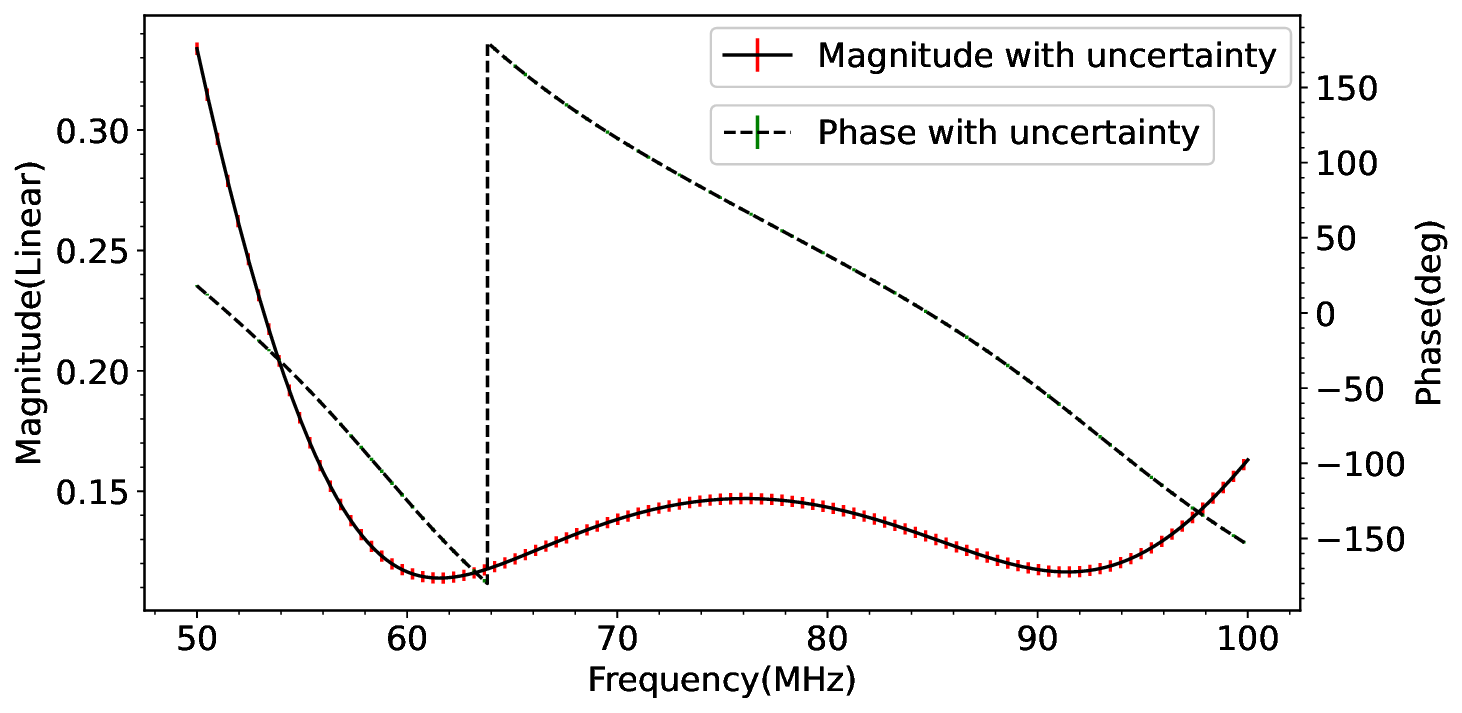}
\includegraphics[width=0.7\columnwidth]{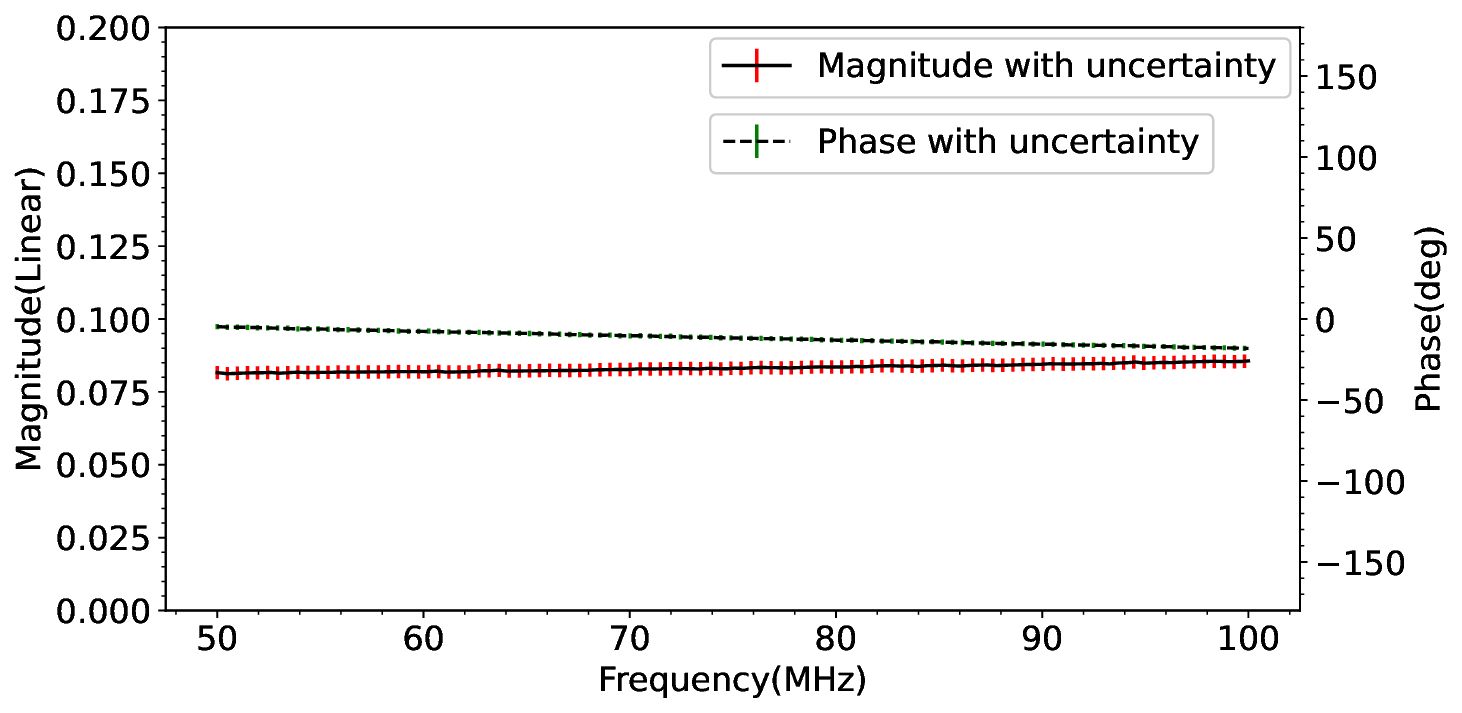}
\includegraphics[width=0.7\columnwidth]{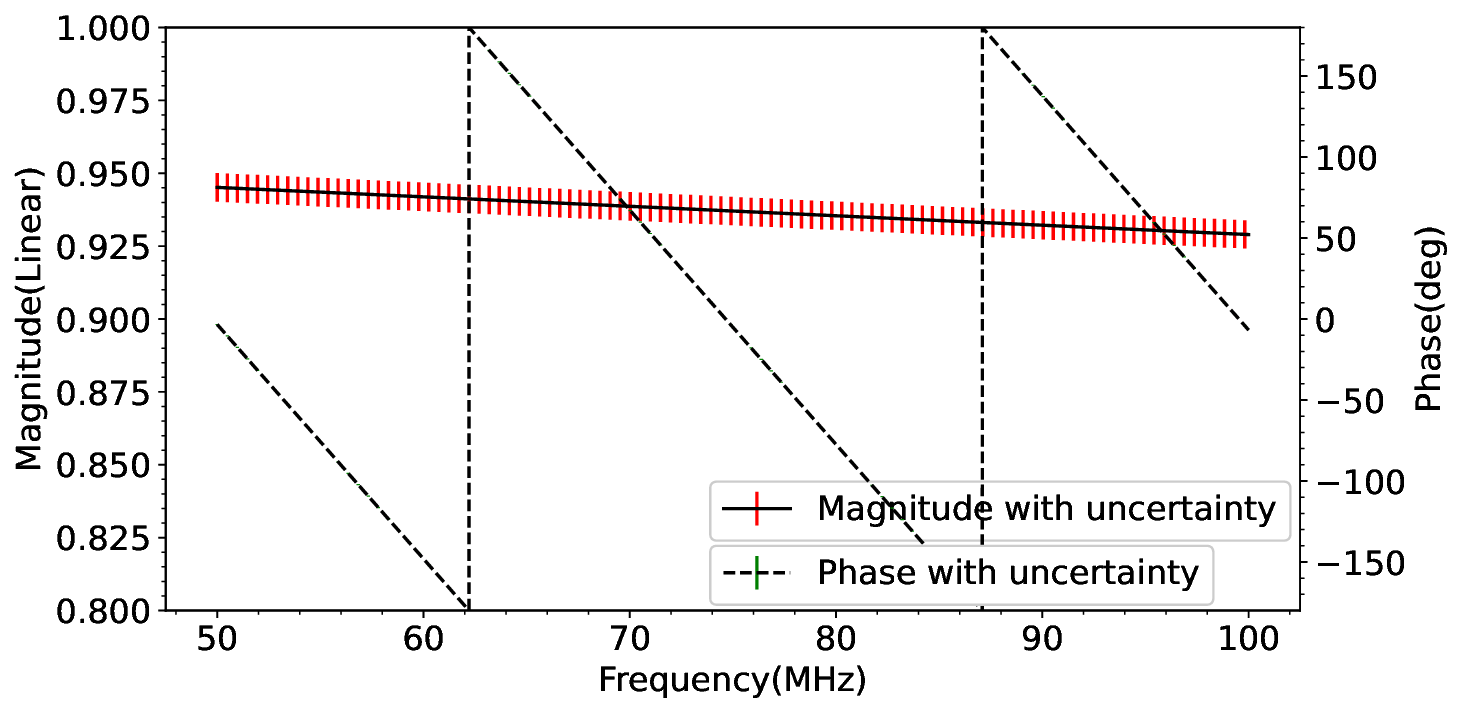}
\caption{\textbf{Top}: Reflection coefficients of the antenna (CST simulation). \textbf{Middle}: Reflection coefficients of the LNA receiver (measurement result of a custom-made LNA module. The LNA consists of two Wantcominc WHM0003AE \citep{wantcominc} transistors). \textbf{Bottom}: Reflection coefficients of a 5 m open cable (cable model as given by Equation~(\ref{eq:gamma open})). The solid curves show magnitude, dashed curves show phase, and error bars show VNA measurement uncertainty (discussed below in Section \ref{sec:Simulation with VNA uncertainty}) and Appendix~\ref{appb}.}
    \label{fig6}
\end{figure}

Figure~\ref{fig5} shows the calculated magnitude and phase uncertainties of the reflection coefficient measurement for the specific VNA and the calibrator standard set. Below, we use these results as the VNA measurement error in our simulation. The distribution of the systematic error is usually unknown, and the uniform distribution is assumed. The standard deviation is calculated as 
$({\rm maximum~value})/\sqrt{3}$, and the total uncertainty is estimated by taking the root sum square (RSS) of the systematic and random errors, although the errors introduced through cable movement or temperature drift are not included. 
As shown in the figure, the uncertainty in the phase is large when the magnitude of the reflection coefficient is small, because in that case, the signal is too weak to be measured precisely. Figure~\ref{fig6} shows the reflection coefficients of the antenna, the LNA receiver, and the open cable with measurement errors.

%%%%%%%%%%%%%%%%%%%%%%%%%%%%%%%%%%%%%%%%%%

\section{Error Propagation}
\label{sec:Error propagation}

We now consider how the errors affect the measured signal. We first consider the errors arising from the receiving system, which we call receiver errors, and then those induced by the VNA measurement errors, which we call VNA errors. We use the reconstruction simulation pipeline to simulate and analyze the impact of each error. We also consider the application of attenuator in the system, which, in some cases, could reduce the system error.

%%%%%%%%%%%%%%%%%%%%%%%%%%%%%%%%%%%%%%%%%%%%%%%%%%%%%%%%%%%
\subsection{Receiver Error Simulation}
\label{sec:Receiver error simulation}

\begin{figure}[H] 
%\centering
	\includegraphics[width=0.98\columnwidth]{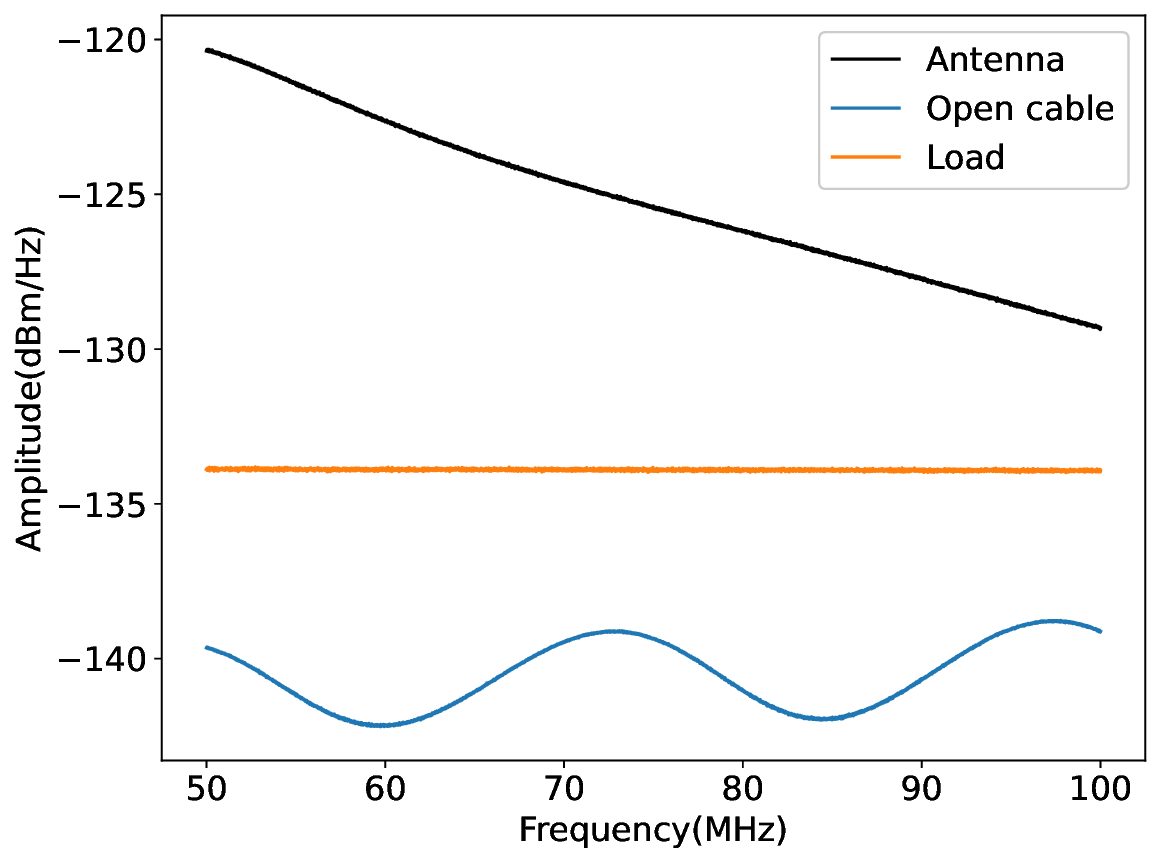}
    \caption{Simulated data of the global spectrum measurement system for an integration period of 10 s. The black, orange, and blue curves are for the input switched to antenna, load, and the  open cable,~respectively. 
    }
    \label{fig7}
\end{figure}

The noise figure of the receiver in our example system is approximately 1 dB in the observation band (c.f. Figure~\ref{fig4}), which corresponds to a 75 K receiver noise temperature for a load physical temperature of 290 K, while the transmission line loss between the antenna and the receiver is negligible. The input signal is amplified by the receiver and then fed into a digital data acquisition system; here, we neglect the errors in the digital system. Inspired by actual experimental results, we model the noise wave parameters $T_u, T_c,$ and $T_s$ as linear functions of frequency, $T_u=0.04 \nu_{\rm MHz} +31~\rm K$, $T_c=0.04~\nu_{\rm MHz} +6~\rm K$, and $T_s=0.06~\nu_{\rm MHz} +6 \rm~K$, where $\nu_{\rm MHz}$ is the frequency in units of MHz.
We assume the sampling rate is 100 MSPS for 50 MHz bandwidth, and the FFT length is 16,384, to produce 8192 output spectral bins. A total of 500,000 s (138.89 h) of mock data for antenna input and the same amount of mock data for the load and open cable inputs are generated. Thermal noise is also added in each terminator, with $\delta T = (T+T_{\rm rec})/\sqrt{\Delta \nu \tau}$, where $\tau$ is integration time (set to 10 s), $\Delta \nu$ is bandwidth, $T_{\rm rec}$ is receiver noise temperature, and $T$ is the temperature of the terminator, i.e., when switched to antenna $T=T_{\rm sky}$; when switched to open cable or load, $T$ equals the ambient temperature. The raw spectrum accumulated from each input is averaged, which reduces random noise.
Figure~\ref{fig7} shows the simulated receiver output for the three input sources with an integration period of 10 s. The data for the open cable show an oscillation with respect to the frequency, as the reflection at the end of the cable sets up a standing wave in the system. 
If the reflection coefficients are known, we can then derive the noise wave parameters $T_u, T_c,$ and $T_s$ as a function of frequency by the least square fitting of the calibration data. 

\begin{figure}[H]

	\includegraphics[width=1\columnwidth]{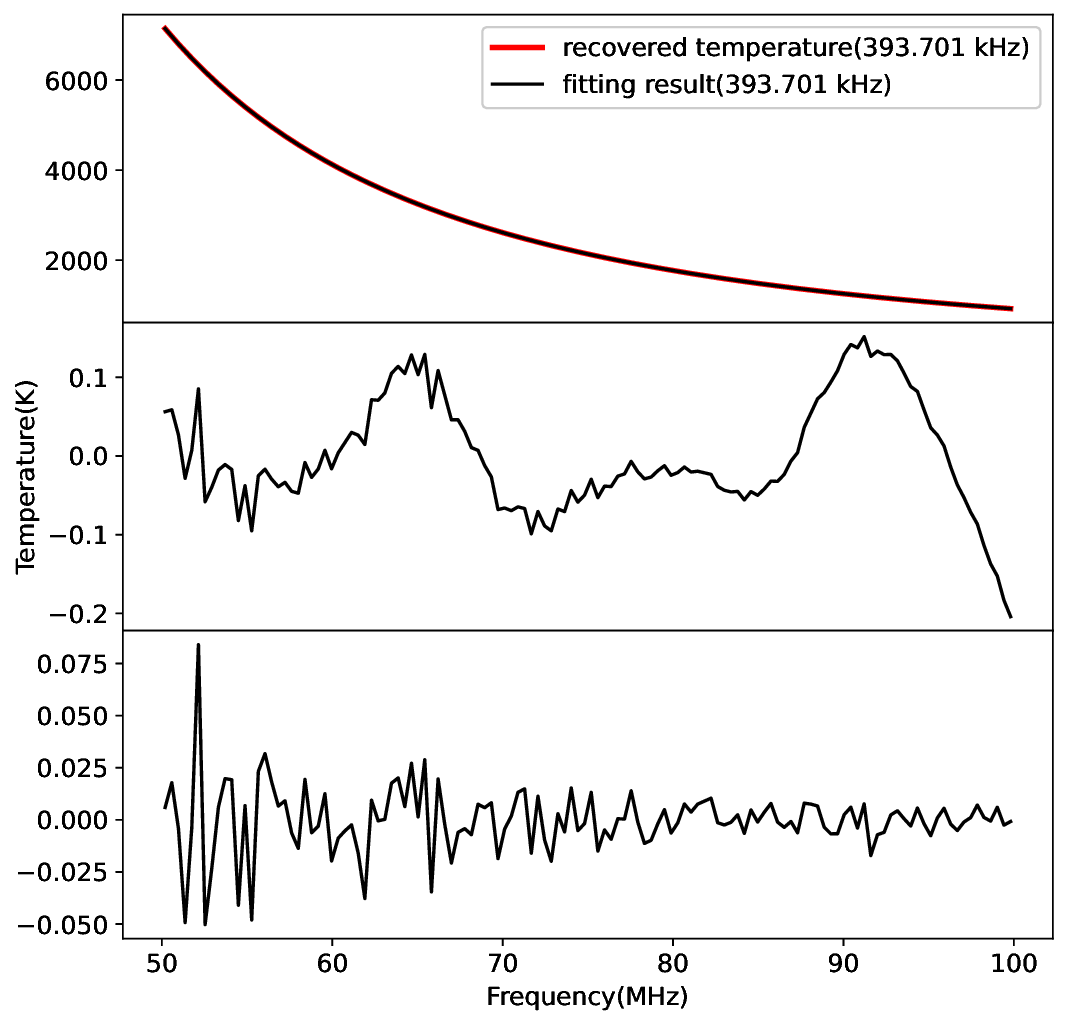}
    \caption{Top: The reconstructed sky spectrum (smoothed) and its fitting with the foreground model. Middle: residuals of the fit. Bottom: residuals of fitting reconstructed sky spectrum with both foreground and 21 cm model simultaneously.}
    \label{fig:recovered integration sky temperature}
\end{figure}

\begin{figure}[H]

\begin{adjustwidth}{-\extralength}{0cm}
\centering 
\includegraphics[width=1.3\columnwidth]{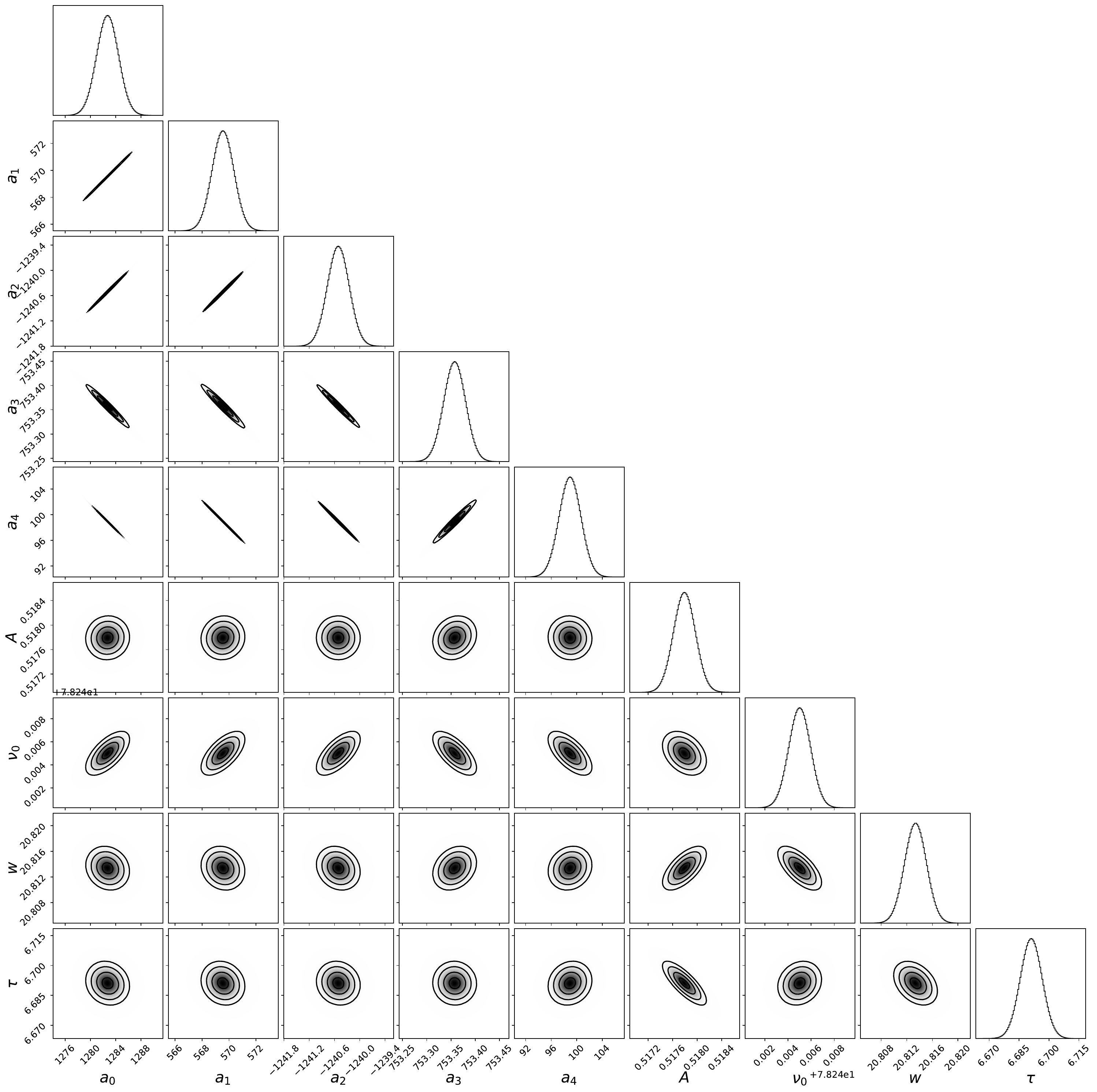}
\end{adjustwidth}
    \caption{{Distributions for the foreground and 21 cm model parameters.}}
    \label{fig:Likelihood distributions} 
\end{figure}

We plot the calibrated sky spectrum in the top panel of Figure~\ref{fig:recovered integration sky temperature}.  To reduce scatter, the data of every 64 frequency channels are rebinned to one; the resulting spectrum has 128 frequency bins with a width of 393.7 kHz. This is fitted with the five term foreground model given by Equation~(\ref{eq:foreground}) in the middle panel, and the foreground plus 21 cm signal (Equation~ (\ref{eq:21 cm absorption})) model in the bottom panel. As expected, in the foreground-only model, the residue appears to have spectral features, while the foreground+21 cm model yields flat residues. Figure~\ref{fig:Likelihood distributions} shows the distributions of the foreground model parameters ($a_0$, $a_1$, $a_2$, $a_3$, $a_4$) and the 21 cm model parameters ($A$, $\upsilon_0$, $w$, $\tau$) derived from the mock observation. Although there is some degeneracy in the foreground polynomial coefficients, the 21 cm model parameters are basically uncorrelated among themselves.

In the above, we have assumed that the system is time-independent, ignoring the time variation in the system characteristics. However, the gain of the system drifts over time, especially if there are temperature variations in the circuit. If the overall gain drifts $\pm$0.5 dB, while keeping other system parameters stable and bandpass shape unchanged,  the simulations show that it causes approximately 20 mK deviation in the results for our system. The gain could be relatively calibrated with a noise source (usually a diode) relatively and absolutely with a thermal load. The calibration error is given by 
$\delta g = \delta T_{\rm cal}/T_{\rm cal}$. If we assume the noise source temperature fluctuation is given by  
$\delta T_{\rm cal} = T_{\rm cal}/\sqrt{\Delta \nu \tau}$, then for an integration time of 10 s and bandwidth of 393.7 kHz, we have
$\delta g = 1/\sqrt{\Delta \nu \tau} \sim 5 \times 10^{-3} $ or 0.002 dB.  
 
\begin{figure}[H]

    \includegraphics[width=0.8\columnwidth]{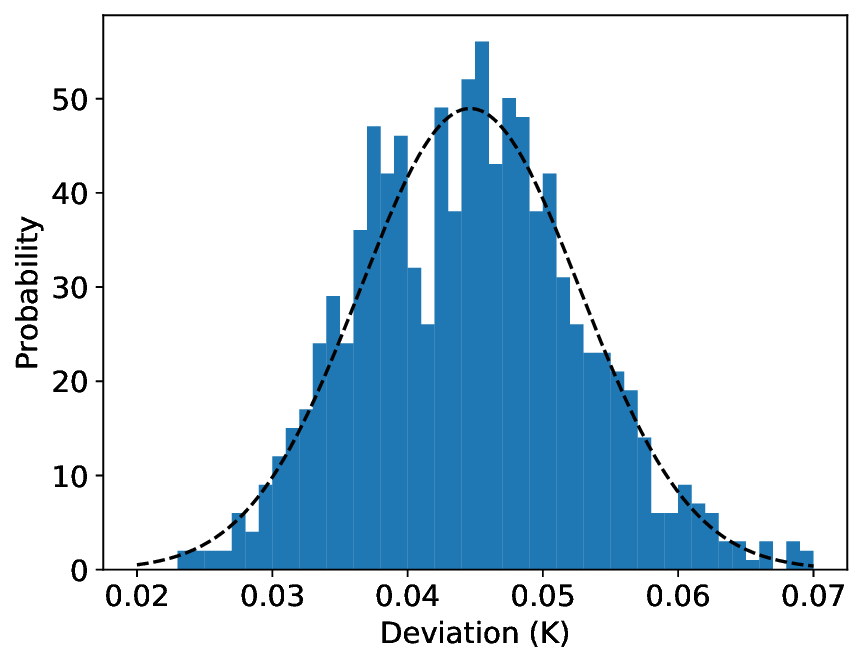}
    \caption{Probability distribution of recovered residue deviation from expected signal. Statistics show the error with mean value of 45 mK, standard deviation of 0.008.}
    \label{fig:deviation probability with load thermal noise}
\end{figure}

 \newpage
The output of the load in the circuit also has thermal fluctuations. This is used to calibrate the $T_0$ by applying Equation~(\ref{eq:load temperature approx}) so the fluctuation would induce an error of the size $\delta T_0 \sim T_{\rm amb}/\sqrt{\Delta \nu \tau}$. We set the integration time to 10 s, as before, while we use three cycle's thermal load data (30 s) for receiver calibration. This induces a variation in the calibrated gain of the system. Figure~\ref{fig:deviation probability with load thermal noise} shows the probability distribution of the deviation from the true signal in our simulation, indicating an error with a mean value of 45 mK. This error could be further reduced if a longer integration time is used. An optimal time scale could be achieved by equalizing the thermal noise and the error induced by the drift; the latter depends on the stability of the system.  

\subsection{VNA Error Simulation}
\label{sec:Simulation with VNA uncertainty}

\subsubsection{{Error Propagation in Sky Temperature}}

The system calibration depends on the VNA measurement results. However, even with the finest test equipment, there are still residue errors, as shown in {Figure}~\ref{fig5}, which could bias the result. We now simulate the impact of the VNA measurement. The model is the same as above but without thermal noise. 

For the reflection coefficients of the antenna, the load, and the open cable, we assume the VNA measurements have errors, as given by the VNA specification and shown above in {Figure}~\ref{fig5}.  However, such specifications are often not given in terms of standard deviations; to be on the safe side, the uniform distribution is assumed. The standard deviation is calculated as $({\rm maximum~value})/\sqrt{3}$. Such error is also likely correlated over the frequencies, although the exact variation is unknown. 
We have plotted the magnitude and phase of the antenna, receiver, and the open cable reflection coefficients together with their error bars
in {Figure}~\ref{fig6}, computed as prescribed in Section~\ref{sec:models and simulation methods}. In each panel, the solid and dashed curves show the magnitude (read number from vertical axis on the left) and phase (read number from vertical axis on the right), respectively, and the uncertainty is shown in the error bars. Note here the scales are different for each panel, as in each case, the variation with frequency is different.

The sky temperature can be derived from the observed quantities as:
\begin{equation}
    \begin{aligned}
        T_{\rm sky}=&\frac{T_{\rm ant}-T_0}{(1-\lvert \Gamma_{\rm ant} \rvert^2)\lvert F_{\rm ant} \rvert^2}-\frac{T_u\lvert \Gamma_{\rm ant} \rvert^2}{1-\lvert \Gamma_{\rm ant} \rvert^2}-
        \frac{(T_c \cos(\phi)+T_s \sin(\phi))\lvert \Gamma_{\rm ant} \rvert}{(1-\lvert \Gamma_{\rm ant} \rvert^2)\lvert F_{\rm ant} \rvert}
        \label{eq:Tsky1}
    \end{aligned}
\end{equation}

The errors in each quantity propagate into the sky temperature and affect the reconstruction of the 21 cm temperature. We can use simulations to illustrate the magnitude and pattern of each source of error. 

The error induced in the sky temperature by measurement errors can be estimated by the error propagation formula:
\begin{equation}
    \begin{aligned}
	\delta T_{\rm sky}=\frac{\partial T_{\rm sky}}{\partial \Gamma_{\rm ant}}\delta \Gamma_{\rm ant} + 
	\frac{\partial T_{\rm sky}}{\partial \Gamma_{\rm rec}}\delta \Gamma_{\rm rec} +
	\frac{\partial T_{\rm sky}}{\partial \Gamma_{\rm open}}\delta \Gamma_{\rm open}
        \label{eq:calculating delta sky temperature }
    \end{aligned}
\end{equation}
where $\delta \Gamma_{\rm ant}$, $\delta \Gamma_{\rm rec}$, and $\delta \Gamma_{\rm open}$ denotes the VNA measurement error 
on the antenna, receiver, and open cable,  respectively.  $\Gamma_{\rm ant}$ is
present in the expression of sky temperature Equation~~(\ref{eq:Tsky1})  (also through the $F$ term, c.f. Equation~~(\ref{eq:F term})) so the derivative can 
be computed directly. The sky temperature 
is not directly dependent on $\Gamma_{\rm open}$, but $T_u, T_c,$ and $T_s$ are derived from the calibration measurement given in Equation~(\ref{eq:open temperature}); therefore,
\begin{equation}
    \begin{aligned}
      \frac{\partial T_{\rm sky}}{\partial \Gamma_{\rm open}}=  \frac{\partial T_{\rm sky}}{\partial T_u} \frac{\partial  T_u}{\partial \Gamma_{\rm open}}
      + \frac{\partial T_{\rm sky}}{\partial T_c} \frac{\partial  T_c}{\partial \Gamma_{\rm open}} +  \frac{\partial T_{\rm sky}}{\partial T_s} \frac{\partial  T_s}{\partial \Gamma_{\rm open}}
        \label{eq:Tsky_open}
    \end{aligned}
\end{equation}

As $T_u, T_c,$ and $T_s$ are derived from a numerical fitting procedure of the measured $T_{\rm open} (\nu)$ over a frequency range, we obtain the 
 partial derivatives numerically. Finally, $\Gamma_{\rm rec}$ is present in the sky expression directly through $F_{\rm ant}$  (Equation~~(\ref{eq:F term}))
 and indirectly by affecting $T_u, T_c,$ and $T_s$ (Equation~(\ref{eq:open temperature})) and $T_0$ (through Equation~(\ref{eq:load temperature approx})). Therefore, we have 
\begin{eqnarray}
      \frac{\partial T_{\rm sky}}{\partial \Gamma_{\rm rec}} &=&  \left. \frac{\partial T_{\rm sky}}{\partial \Gamma_{\rm rec}} \right \lvert_{\rm direct}
      + \frac{\partial T_{\rm sky}}{\partial T_u} \frac{\partial  T_u}{\partial \Gamma_{\rm rec}}
      + \frac{\partial T_{\rm sky}}{\partial T_c} \frac{\partial  T_c}{\partial \Gamma_{\rm rec}} 
      +  \frac{\partial T_{\rm sky}}{\partial T_s} \frac{\partial  T_s}{\partial \Gamma_{\rm rec}}+\frac{\partial T_{\rm sky}}{\partial T_0} \frac{\partial  T_0}{\partial \Gamma_{\rm rec}} 
        \label{eq:Tsky_load}
\end{eqnarray}
where the first term on the R.H.S. of the equation is obtained by calculating the partial derivative using the explicit function form directly, while 
the remaining terms are calculated in a way similar to the $\Gamma_{\rm open}$ case.

We estimate the VNA measurement errors according to the value of reflection coefficients for the antenna, receiver, and open cable as shown. The error in magnitude is at the level of a few times $ 10^{-3}$, and the error in the phase is at the 
level of $1.5^\circ$. In the frequency range of 50--100 MHz, these errors only have slight variations. 

\begin{figure}[H]

    \includegraphics[width=0.8\columnwidth]{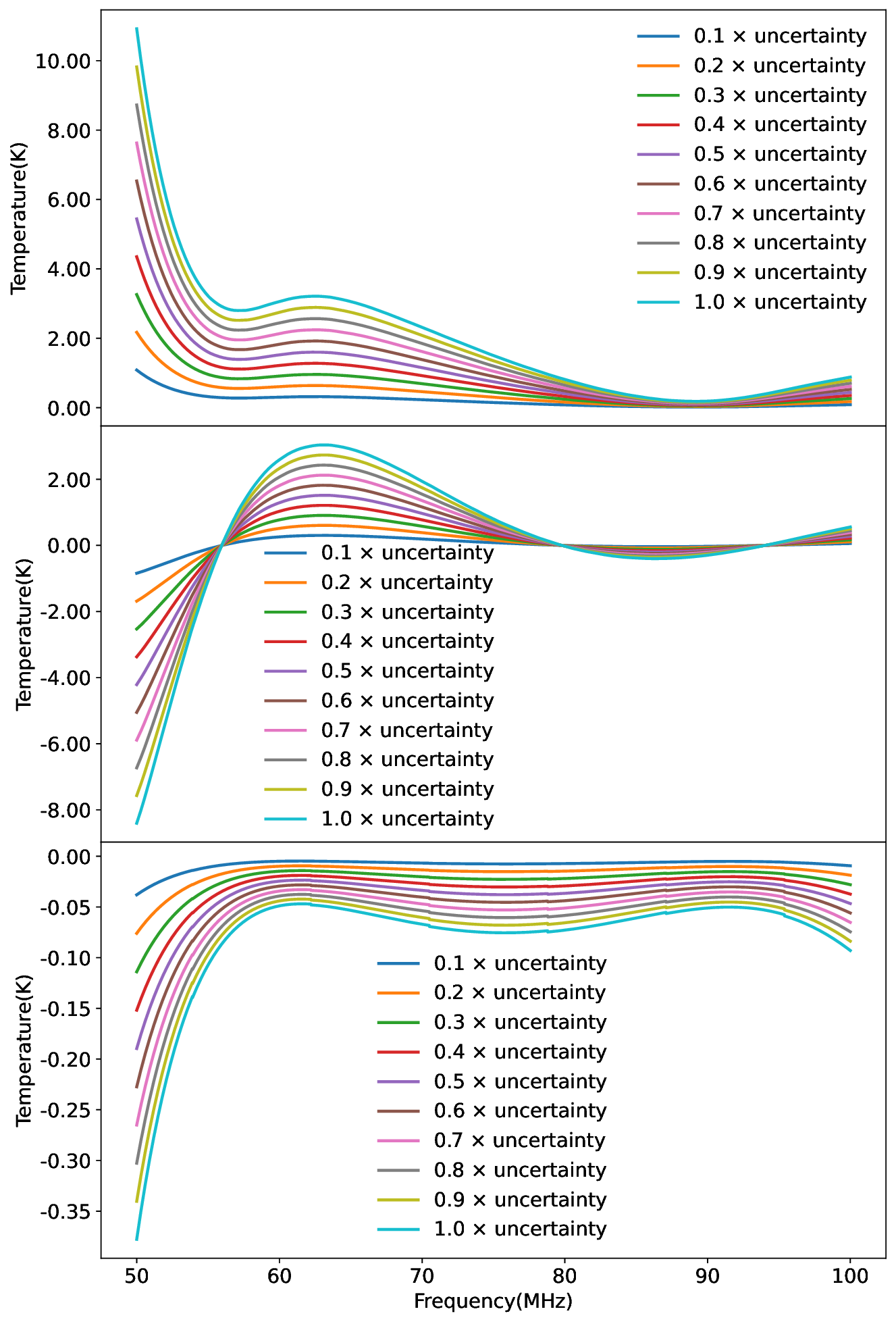}
    \caption{Top (\textbf{a}): Sky temperature deviation with VNA magnitude error for the antenna measurement only. Middle (\textbf{b}): Sky temperature deviation with VNA magnitude error for the receiver measurement only. Bottom (\textbf{c}): Sky temperature deviation with VNA magnitude error for the open cable measurement only.}
    \label{fig:computed and recovered delta tsky}

\end{figure}

In Figure~\ref{fig:computed and recovered delta tsky}a--c, we plot separately the deviation in reconstructed sky temperature caused by the VNA error when it is used to measure the reflection coefficients of the antenna, receiver, and open cable, respectively. As the errors given in the manual of VNA are $({\rm maximum~values})/\sqrt{3}$ and are perhaps correlated over the frequencies, we plot the induced errors with the error set in the range of 0.1 to 1.0 times the nominal uncertainty, which varies smoothly over the frequencies, and the multiple curves remind us the possible size of the error. From this figure, we see the induced errors are large for the antenna and receiver measurements,  while the open cable measurement error has less impact on results. 

In Figure~\ref{fig:deltaT21 cm}a, we plot the recovered sky temperature deviation with the VNA measurement error on antenna, receiver, and open cable simultaneously---this is closer to the real case as the same VNAs are switched to make these measurements. The results show that for an error equal to 10\% (0.1) of the nominal error, there is a deviation of approximately 0.5 K in reconstructed sky temperature, while for the nominal value, the deviation is approximately 6 K in sky temperature at 63 MHz.

\subsubsection{{Error Propagation in 21 cm Signal}}

\begin{figure} [H]
	\subfloat[\centering] {\includegraphics[width=0.8\columnwidth]{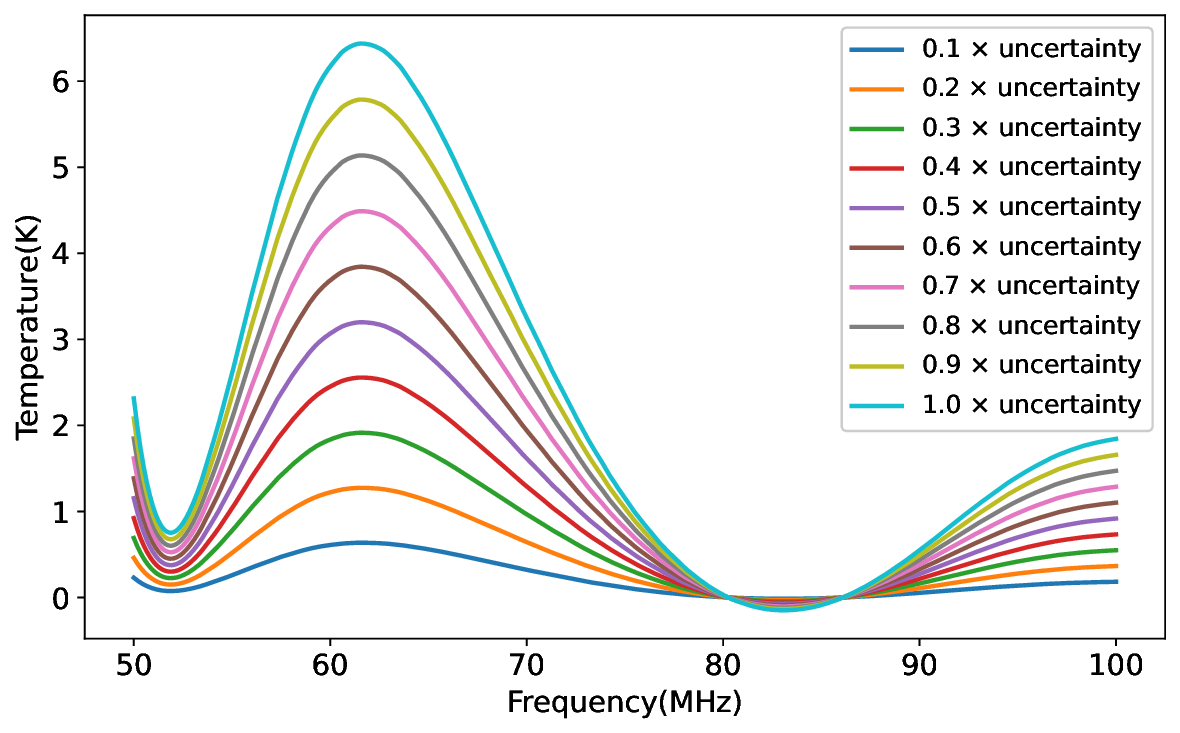}}
	\vspace*{2pt}
	\subfloat[\centering] {\includegraphics[width=0.8\columnwidth]{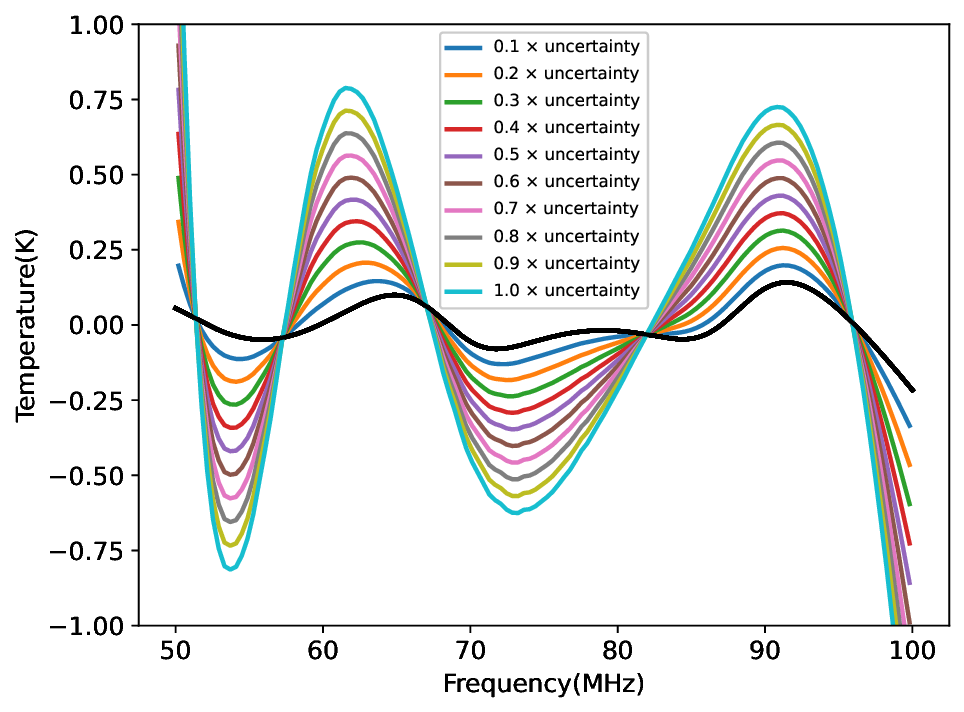}}
	\caption{Top (\textbf{a}): Sky temperature deviation with VNA magnitude error on all measurements. Bottom (\textbf{b}): The reconstructed 21 cm signal deviation.}
	\label{fig:deltaT21 cm} 
\end{figure}

The sky temperature spectrum is fitted with the foreground and 21 cm model spectra, so the error in the reconstructed 21 cm signal is a nonlinear function of the VNA measurement error, which does not have an analytical expression. However, one can apply the fitting algorithm to the sky temperature spectrum with the error and then derive the residue with 21 cm signal reconstruction. We plot in Figure~\ref{fig:deltaT21 cm}b the 21 cm reconstruction results. The error in the reconstructed 21 cm signal is not directly proportional to the whole deviation in sky temperature spectrum, as the smooth components are removed as foregrounds. The residue deviations in the 21 cm spectrum have much smaller magnitude than the whole sky temperature spectrum shown in Figure~\ref{fig:deltaT21 cm}a, but they are less smooth, showing more oscillatory features, and the deviation magnitude also increases as the VNA error grows. The deviation level can reach $\pm\sim750$ mK maximum in this model. 
We also calculated the RMS deviation value of the reconstructed 21 cm signal for each magnitude uncertainty level, e.g., for half (0.5) the maximum uncertainty in magnitude, the mean uncertainties of the reflection coefficient of the antenna, the receiver, and the open cable over the 50--100 MHz band are approximately 0.001, 0.001, and 0.002  (linear), respectively, and the RMS deviation value of the reconstructed 21 cm signal is 200 mK.

\begin{figure}[H]

\includegraphics[width=0.8\columnwidth]{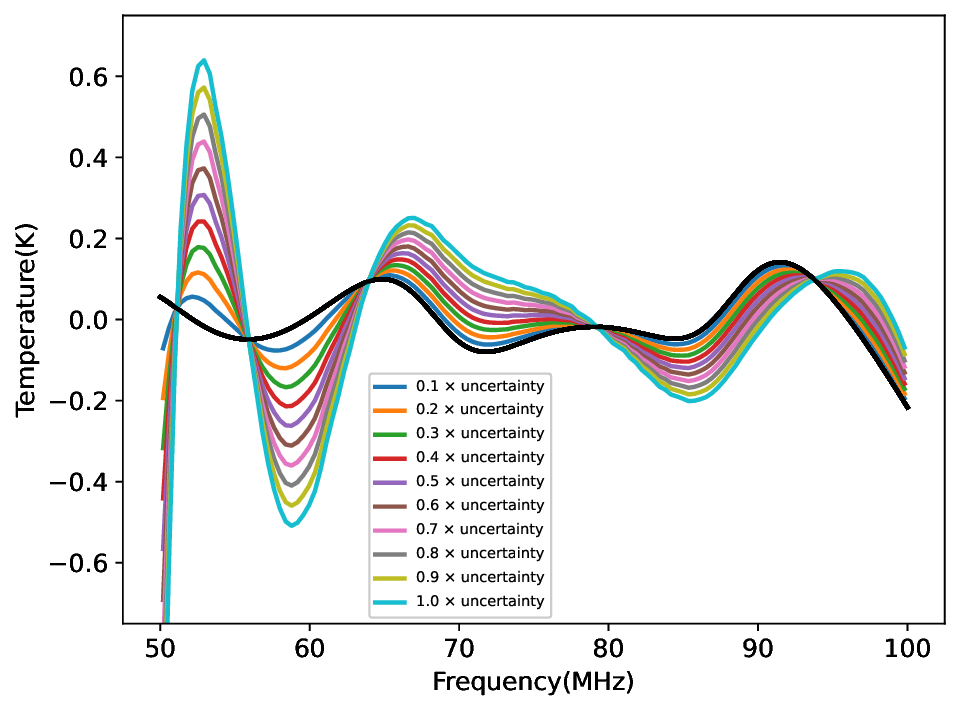}
    \caption{The reconstructed 21 cm spectrum residuals with VNA phase measurement error. Solid curve represents the residual model, and different colored curves represent recovered residuals with different phase measurement errors, error from 0.1 to 1.0 times the nominal phase error. }
    \label{fig:error in pha constant}
\end{figure}

The global spectrum experiment considered in the present paper measures the total received power from one antenna; the phase information of the wave is not recorded. Despite this, the phase error in the reflection coefficient also affects the total power spectrum reconstruction.
In the above, we have considered the error induced by the magnitude of the VNA measurement. Next, we consider the effect of the error in the VNA phase measurement. The phase measurement error also depends on the reflection coefficient. As shown in {Figure}~\ref{fig5}, for a reflection coefficient larger than 0.4 ($S_{11}=-7.95\rm~dB$), the phase uncertainty curve is quite flat. In the simulation, we set a series of 10 relative error values; for example, the uncertainty for open cable ranges from the relatively small $0.03^\circ$ to the maximum uncertainty of $0.3^\circ$, while for the receiver, the uncertainty ranges from the relatively small $0.15^\circ$ to the maximum uncertainty of $1.5^\circ$. Figure~\ref{fig:error in pha constant} shows the reconstruction results. As the error grows, the deviation become greater, while the absorption shape of the reconstruction result is similar to the residual model. We also calculated the RMS deviation value of the reconstructed 21 cm signal for each phase uncertainty level, e.g., for a half (0.5) of the maximum phase uncertainty, the mean uncertainties of the antenna, the receiver, and the open cable over the 50--100 MHz band are approximately $0.48^\circ$, $0.78^\circ$, and $0.15^\circ$, respectively, and the RMS deviation value of the reconstructed 21 cm signal is 40 mK.

\begin{figure}[H]

	\includegraphics[width=0.8\columnwidth]{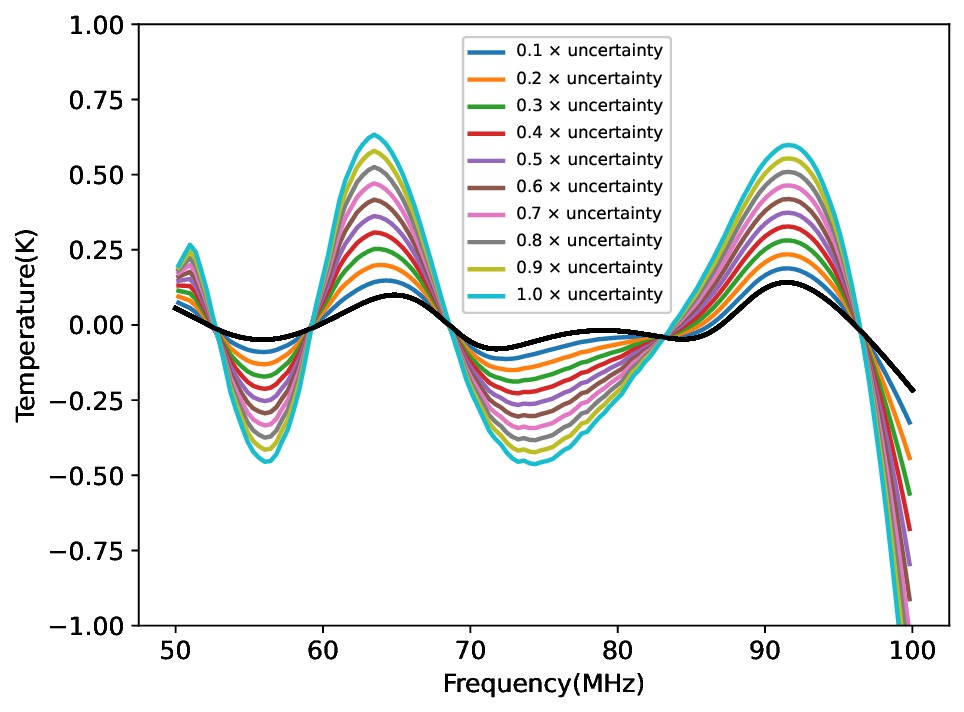}
    \caption{Recovered residuals with both VNA magnitude and phase measurement errors using a 15 m open cable for solving noise wave parameters. Both magnitude error and phase error are added. The magnitude error is from 0.1 to 1.0 times the nominal magnitude error, accompanied with phase error from 0.1 to 1.0 times the nominal phase error. Black curve represents the case without error, different colored curves represent different measurement errors.}
    \label{fig:recovered residuals with 15m cable}
\end{figure}

To evaluate whether the length of the calibrator open cable has an impact on the calibration accuracy and 21 cm signal reconstruction when the VNA measurement errors are added in, we also 
simulated the cases with open cables of length from 3 m to 30 m. The result for the case of a 15 m open cable is plotted in Figure~\ref{fig:recovered residuals with 15m cable}. The errors on the magnitude and phase are added to the terminators' reflection characteristics, then the same simulation and reconstruction procedures are applied as before. It seems that there is very little difference in the result, showing that the cable length does not have a significant impact on the precision of the experiment under the premise that the noise wave parameters were solved with high accuracy.

\subsection{Effect of an Attenuator}
\label{sec:Attenuator effect simulation}

In some global spectrum setups, an attenuator is inserted between the antenna and receiver to improve the effective matching between the antenna and the receiver, reducing the standing wave amplitude and alleviating the impact of VNA measurement uncertainty. However, adding a passive device before the LNA also increases the thermal noise of the system. 

\begin{figure}[H]

\includegraphics[width=\columnwidth]{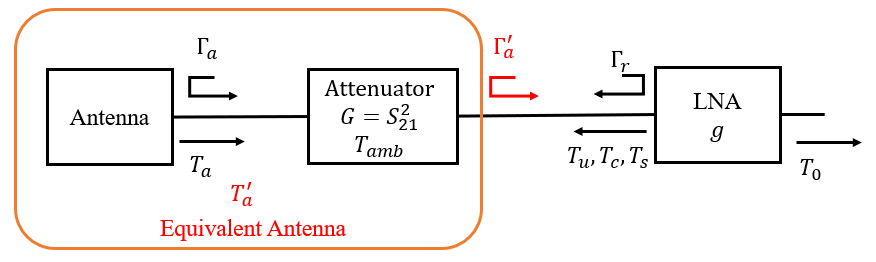}
\caption{The antenna and attenuator forming an equivalent antenna with new parameters. }
\label{fig:equivalent_antenna}
\end{figure}

Below, we treat the antenna terminated with the attenuator as an equivalent antenna, as depicted in Figure~\ref{fig:equivalent_antenna}. In this approach, the parameters associated with the receiver, including $T_u$, $T_c$, and $T_s$ are unchanged, just as when it is replaced by the open cable during calibration. The reflection coefficient of the effective antenna, i.e., the combination of antenna and attenuator is
\begin{equation}
\label{eq:Gamma_ant_att}
\Gamma_{\text{a}}^{\prime}=S_{22}+\frac{S_{12} S_{21} \Gamma_{\text{a}}}{1-S_{11} \Gamma_{\text{a}}}
\end{equation}
where the $S$-parameters are those of the attenuator. For an ideal attenuator, $S_{11} = S_{22} = 0, S_{21} = S_{12}$, and $G=S_{21}^2$, then
\begin{equation}\label{eq:7}
\begin{aligned}
\Gamma_{\text{a}}^{\prime}=G \Gamma_{\text{a}}.
\end{aligned}
\end{equation}

\begin{figure}[H]

\includegraphics[width=\columnwidth]{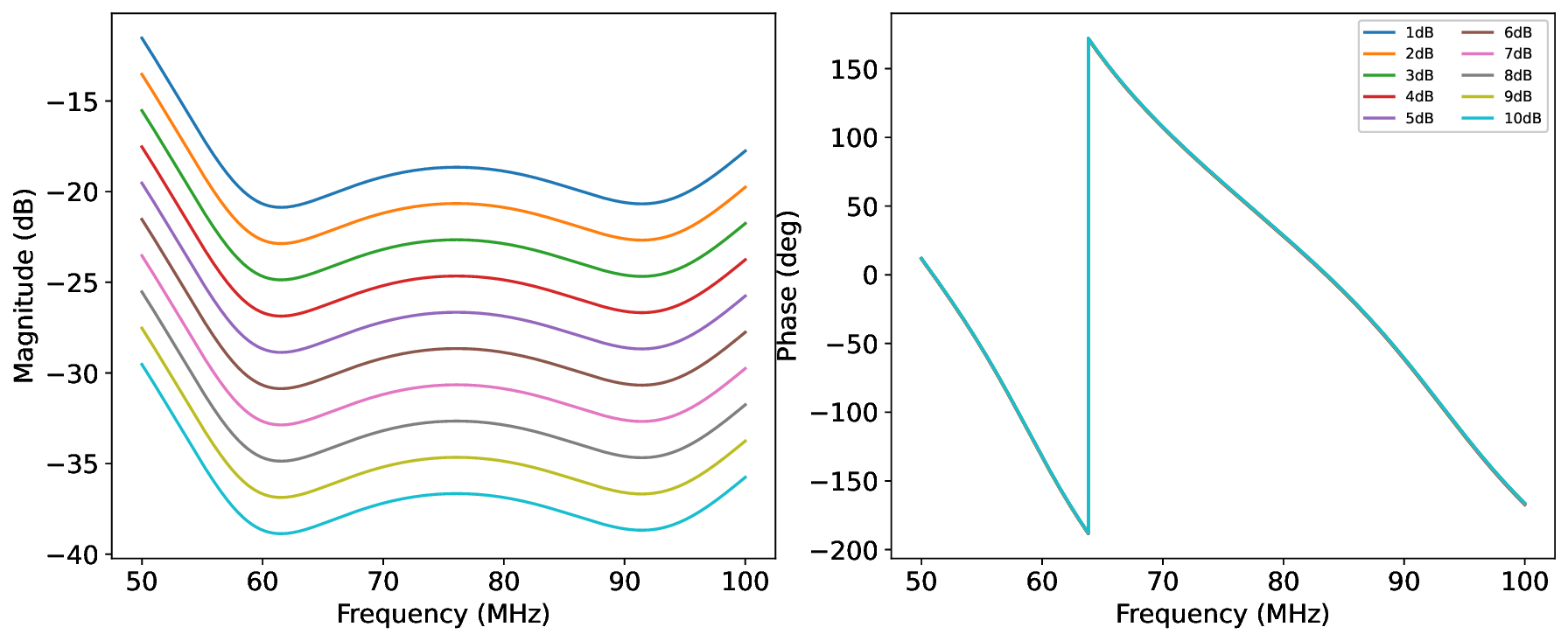}
\caption{The antenna with attenuator model. 
The magnitude and phase are plotted on the left and right, respectively, and the curves are for the attenuation of $1, 2, 3, ..., 10$ dB.}
\label{fig:antenna_receiver_with_attenuator_model}
\end{figure}

Compared with the original reflection coefficients of the antenna, the magnitude of the reflection coefficient is reduced, as shown in Figure~\ref{fig:antenna_receiver_with_attenuator_model}, with an attenuation of $1\sim10$ dB.


\begin{figure}[H]

\includegraphics[width=\columnwidth]{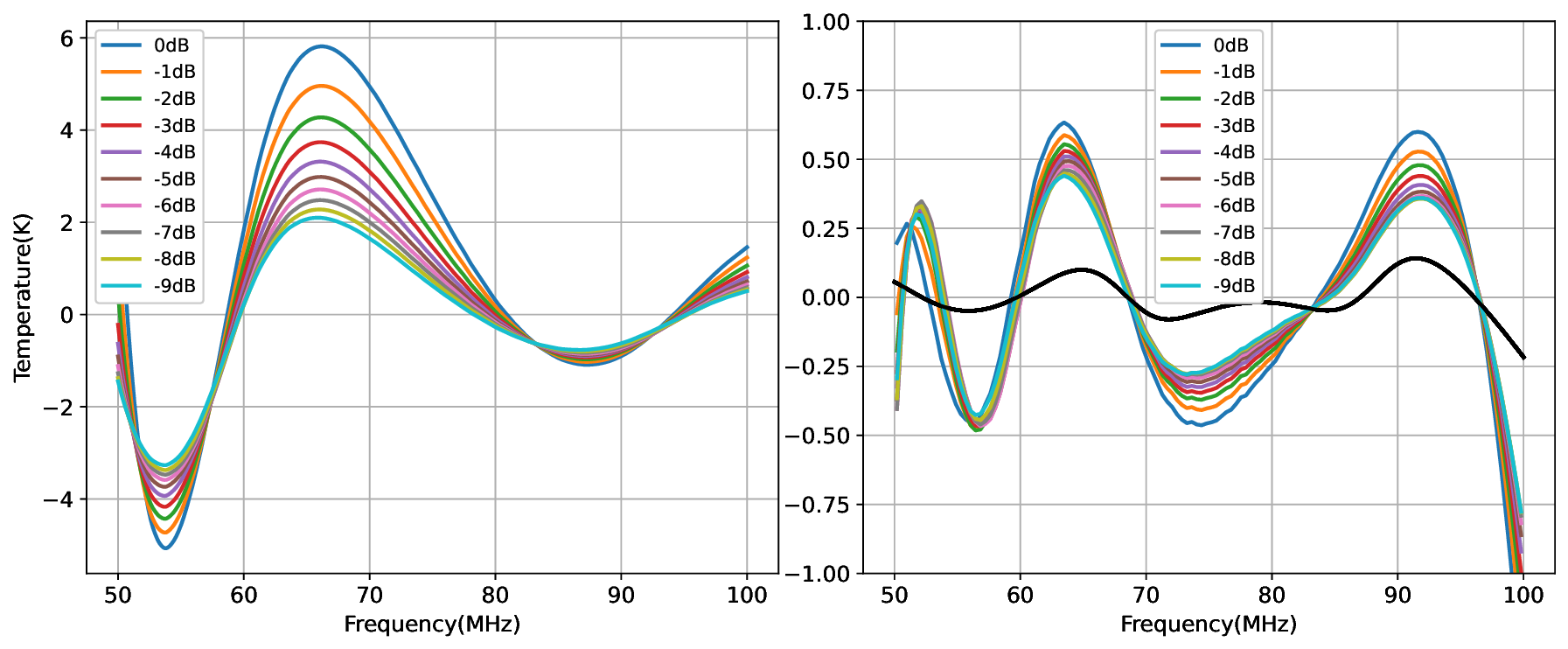}
\caption{Sky temperature and 21 cm signal deviation for the equivalent antenna case. 
}
\label{fig:att_equiv_temp}
\end{figure}

Figure \ref{fig:att_equiv_temp} shows the sky temperature deviation and 21 cm signal deviation for the equivalent antenna and the equivalent receiver model, respectively. Here, we add both magnitude and phase measurement error(1.0$\times$ uncertainty) in the antenna, the receiver, and the open cable. The results show that with the attenuator, the residue error is indeed reduced. For example, the maximum residue is reduced from 0.63 K to 0.44 K for the 21 cm signal for an attenuation of 9 dB.

The original $T_0$ corresponds to the output noise temperature generated by the LNA, which is 75 K for a load with a physical temperature of 290 K. In a global spectrum experiment, the thermal noise in each terminator can be expressed as $\delta T = (T+T_{\rm rec})/\sqrt{\Delta \nu \tau}$, so an increase in the receiver noise temperature results in an increase in thermal noise in the system, which needs a longer integration time to achieve the desired signal-to-noise~ratio.

\begin{figure}[H]
\centering
\vspace{-0.35cm}
\includegraphics[width=0.48\linewidth]{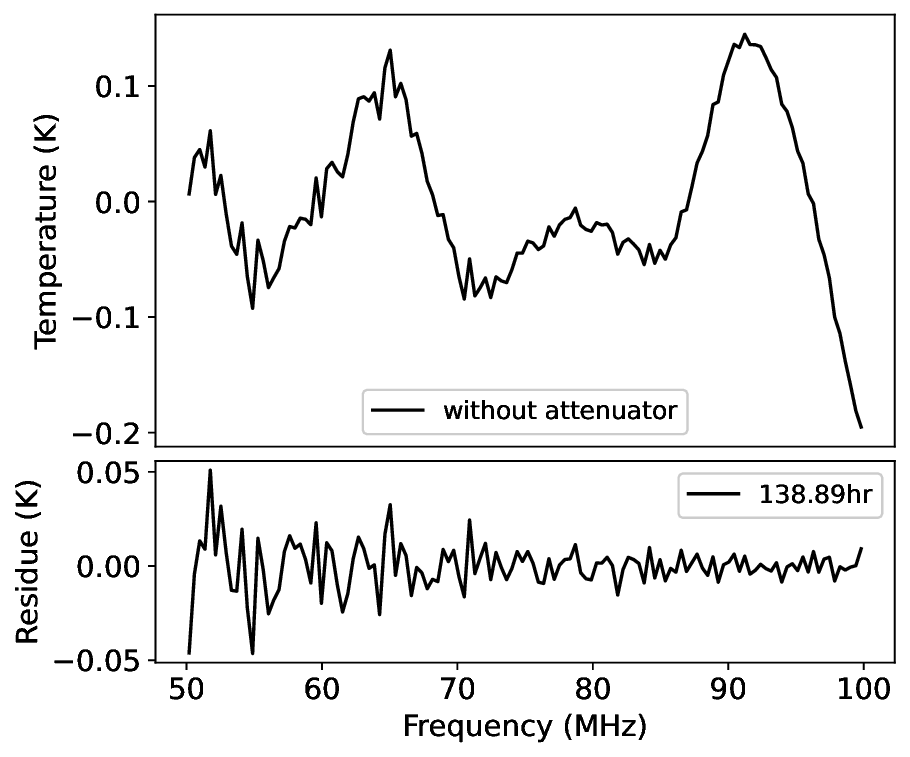}
\includegraphics[width=0.48\linewidth]{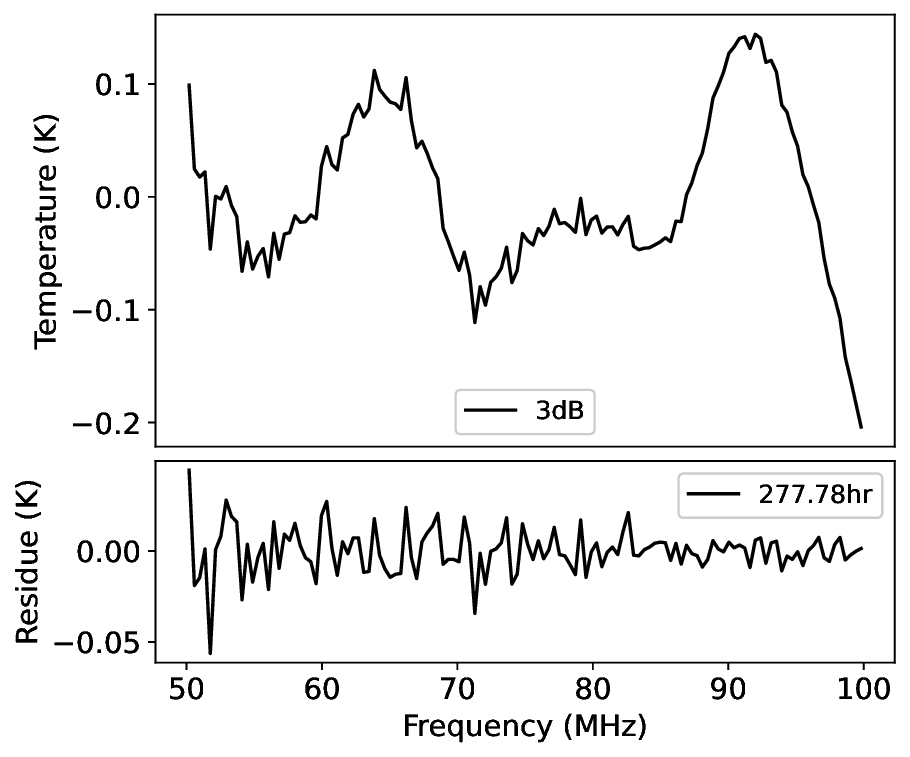}\\
\includegraphics[width=0.48\linewidth]{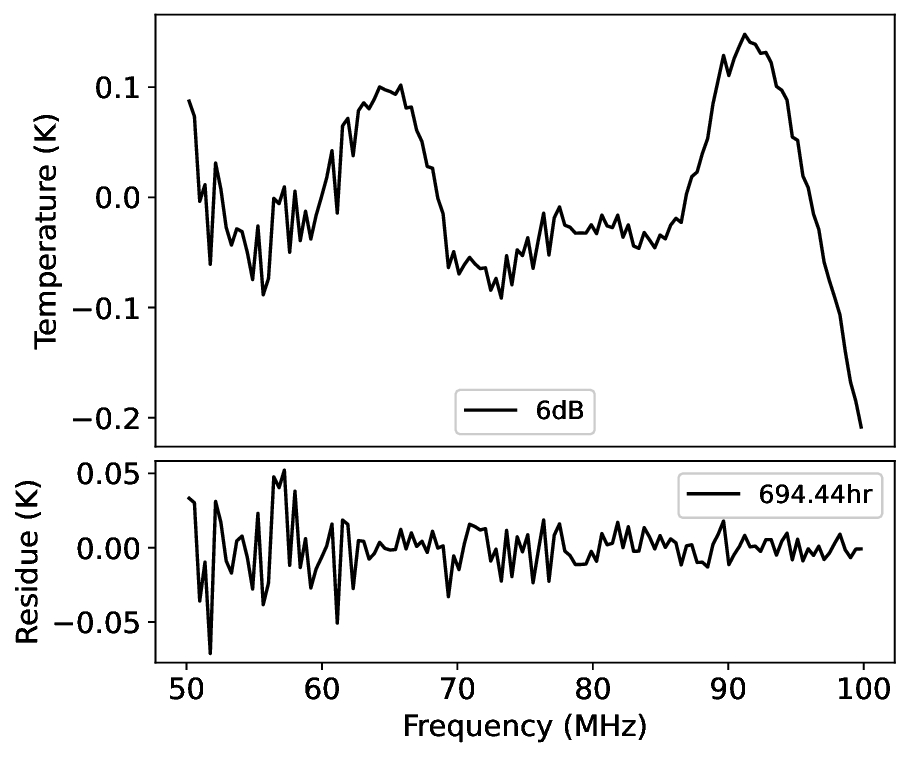}
\includegraphics[width=0.48\linewidth]{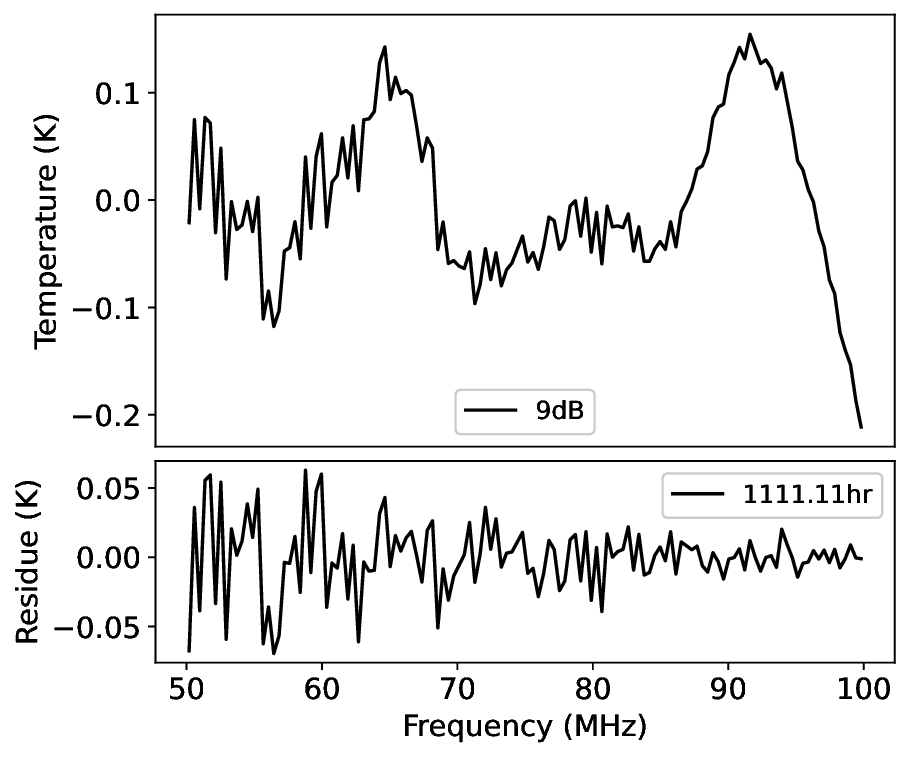}
\caption{Recovered residue with different values of attenuator corresponding with different integration~times.}
\label{fig:recovered residue with different values of attenuator.}
\end{figure}

Figure~\ref{fig:recovered residue with different values of attenuator.} shows the recovered residue after foreground subtraction without the attenuator and cases of 3 dB, 6 dB, and 9 dB attenuation. To keep the RMS of the residue at 0.05 K level, different integration times are required.  The results show that for the 0 dB, 3 dB, 6 dB, and 9 dB attenuator cases, the integration times required are 138.89, 277.78, 694.44, and 1111.11 h, respectively.

From the above analysis, we see the addition of an attenuator between the antenna, and the LNA could reduce the impact of VNA measurement uncertainty but increase the thermal noise. To obtain the best performance, a compromise should be made by choosing a moderate value of attenuation. For the experimental setup considered here, it seems that an attenuation of $\sim$3 dB is near the optimal, whereas adopting a greater attenuation would only reduce the deviation slightly but would require a considerably longer integration time.

\subsection{Discussions}
\label{sec:impacts on absorption feature}

We estimate the induced errors on the final 21 cm signal. The central frequency of the 21 cm signal $\nu_0$ is not influenced much by the pure VNA magnitude error, while the width of the absorption profile $W$ is not influenced much by the pure phase error. However, the typical VNA magnitude measurement errors of 0.001, 0.001, and 0.002 (linear) on the antenna, receiver, and open cable measurements, respectively, could result in a 200 mK deviation in the 21 cm absorption depth $A$ and a 2 MHz deviation on the absorption profile width $W$. The typical phase measurement errors of $0.48^\circ$, $0.78^\circ$, and $0.15^\circ$ on the antenna, receiver, and open cable measurements, respectively, result in a 40 mK deviation on absorption depth and a 2 MHz deviation on the central frequency.

Since $A\sim30(1-\frac{T_R}{T_S})$ mK, where $T_R$ and $T_S$ are the radiation field and spin temperature, suppose $T_S=7$ K which is the adiabatic-cooling temperature at $z\sim17$ for the gas~\citep{Seager1999ApJ,Seager2000ApJS,Wong2008MNRAS,Scott2009MNRAS}, then $\delta A=200$ mK (40 mK) is equivalent to $\delta T_R\sim50$ K (10 K), or $\delta T_S\sim 7$ K (1.5~K). Therefore, the deviation in the 21 cm absorption depth caused by the VNA error, if not properly corrected in the data reduction, might confuse the astrophysical interpretation of the observed 21 cm signal, e.g., whether extra radio background or exotic cooling mechanism is essential and/or whether weak X-ray heating and strong Ly$\alpha$ coupling is~inevitable.

We foresee that techniques could be developed to improve the VNA measurement accuracy so that it would not adversely affect the 21 cm global spectrum experiment. For example, at present, partly, the error given above is for an instrument calibrated with a standard calibration kit, and the parameters for the kit are provided by the manufacturer. More precise measurement of the individual calibrator kit can be made in the laboratory, especially by characterizing and fitting the impedance and the voltage reflection coefficient as functions of frequency. {The error in the phase of the VNA measurement could be reduced by laboratory calibration with an air-dielectric line which has more precisely known delay characteristics \citep{monsalve2016one,ProtheroeUsingAL}}. The VNA error model can also be refined so that a number of parameters could be individually determined by additional calibration measurements \citep{wong2013improving}. The temperature variation of the instrument also could be compensated. In each calibration measurement, beyond what is well understood by the physical model, the residue could also be checked and reduced by empirical methods, e.g., the ripple method \citep{VNAguide}. {Moreover, some novel VNA calibration methods have been developed, with potential possibilities that can be applied to the global spectrum measurement experiments \citep{9139196}.}

\section{Conclusions}
\label{sec:conclusions and future work}

Detecting the 21 cm global spectrum signal from the cosmic dawn and the epoch of reionization is a very challenging experiment, as it requires extremely high measurement precision. In this paper, we present a simulation of the observation and reconstruction of a 21 cm global spectrum experiment, mainly focusing on systematic error propagation and its impact on recovering the 21 cm signal. We model the foreground with a smooth power-law spectrum, and the 21 cm absorption as a negative flattened Gaussian function centered at 78 MHz, as suggested by the EDGES observation result. Our simulation confirms that thermal noise can be reduced to a low level if the system is stable and allows for a reasonably long integration time. Furthermore, a stable and constant deviation in the system gain or receiver noise offset temperature only has a slight impact on the recovered signal. 

Over the broad frequency band needed for the cosmic dawn 21 cm signal measurement, a mismatch in the impedance of the various RF components inevitably generates reflections and standing waves inside the instrument. If not properly accounted for, such features can be confused with the 21 cm signal. Based on the noise wave formulation, this effect can be corrected, but it is necessary to measure the system parameters by performing a high-precision calibration. Since the system calibration accuracy is determined by the VNA measurement results of reflection coefficients of the antenna, the receiver and the calibrator terminators, any measurement error due to the imperfection of the instrument may affect the calibration and the 21 cm reconstruction result. In order to evaluate this possible source of error, we modeled the uncertainty of the reflection coefficient measurement according to the VNA signal flow graph for a typical VNA model and calibrator type with one of the highest accuracy specifications presently available from the industry. The model includes all known possible errors in VNA measurement procedure and in calibration standards, and realistic VNA uncertainty values are used. 
We propagate the error to the deviation in sky temperature and the extracted 21 cm signal from the foreground removal procedure. The results show that even a relatively small error may cause distortion in both the recovered total sky signal and the 21 cm signal, with amplitude comparable with or even larger than the true 21 cm signal, e.g., a typical VNA magnitude measurement error of 0.001, 0.001, or 0.002 (linear) existing in the antenna, the receiver, or the open cable, respectively, would result in a 200 mK deviation on the expected signal. A typical phase measurement error of $0.48^\circ$, $0.78^\circ$, and $0.15^\circ$ in the antenna, the receiver, and the open cable, respectively, would cause a 40 mK deviation. Results also show that an attenuator benefits to the effect that the VNA measurement uncertainties caused. If these kind of systematic errors cannot be reduced in calibration procedures or corrected in the data analysis, they may confuse the astrophysical interpretation of the observed 21 cm signal. It is of vital importance to develop new methods or techniques to further improve the accuracy of the calibration.

\newpage
 
%%%%%%%%%%%%%%%%%%%%%%%%%%%%%%%%%%%%%%%%%%
\vspace{6pt} 

%%%%%%%%%%%%%%%%%%%%%%%%%%%%%%%%%%%%%%%%%%
%% optional
%\supplementary{The following supporting information can be downloaded at:  \linksupplementary{s1}, Figure S1: title; Table S1: title; Video S1: title.}

%%%%%%%%%%%%%%%%%%%%%%%%%%%%%%%%%%%%%%%%%%
\authorcontributions{Conceptualization: E.d.L.A, funding acquisition and supervision: E.d.L.A and X.C.; investigation and data curation: S.S.; methodology: S.S. and F.W.; formal analysis: J.Z. and F.W.; validation: B.Y.; writing---original draft preparation: S.S.; writing---review and editing: X.C. and E.d.L.A.}.
%For research articles with several authors, a short paragraph specifying their individual contributions must be provided. The following statements should be used ``Conceptualization, X.X. and Y.Y.; methodology, X.X.; software, X.X.; validation, X.X., Y.Y. and Z.Z.; formal analysis, X.X.; investigation, X.X.; resources, X.X.; data curation, X.X.; writing---original draft preparation, X.X.; writing---review and editing, X.X.; visualization, X.X.; supervision, X.X.; project administration, X.X.; funding acquisition, Y.Y. All authors have read and agreed to the published version of the manuscript.'', please turn to the  \href{http://img.mdpi.org/data/contributor-role-instruction.pdf}{CRediT taxonomy} for the term explanation. Authorship must be limited to those who have contributed substantially to the work~reported.

\funding{This research was funded by the Scientific Research Instrument and Equipment Development Project of Chinese Academy of Sciences for the global spectrum measurement instrument for cosmic dawn detection (ZDKYYQ20200008), the CAS Strategic Priority Research Program XDA15072106 and CAS grant QYZDJ-SSW-SLH017, the National natural Science Foundation (NSFC) grants 12361141814, 12273070, 12203061. Eloy de Lera Acedo is supported by the Science and Technology Facilities Council (UKRI-STFC, UK) under grant ST/V004425/1. }

\dataavailability{The data underlying this article will be shared on reasonable request to the corresponding authors.}

\acknowledgments{We acknowledge the logistics support provided by the Astronomical Technology Center of NAOC.}

\conflictsofinterest{The authors declare no conflicts of interest.}

%\section*{Appendix A. }
%%%%%%%%%%%%%%%%%%%%%%%%%%%%%%%%%%%%%%%%%%
%% Optional
\appendixtitles{yes} % Leave argument "no" if all appendix headings stay EMPTY (then no dot is printed after "Appendix A"). If the appendix sections contain a heading then change the argument to "yes".
\appendixstart
\appendix
\section[\appendixname~\thesection]{Derivation of Calibration Formula Based on Noise Wave Concept}\label{appa}
%\subsection[\appendixname~\thesubsection]{}

\begin{figure}[H]

	\includegraphics[width=\columnwidth]{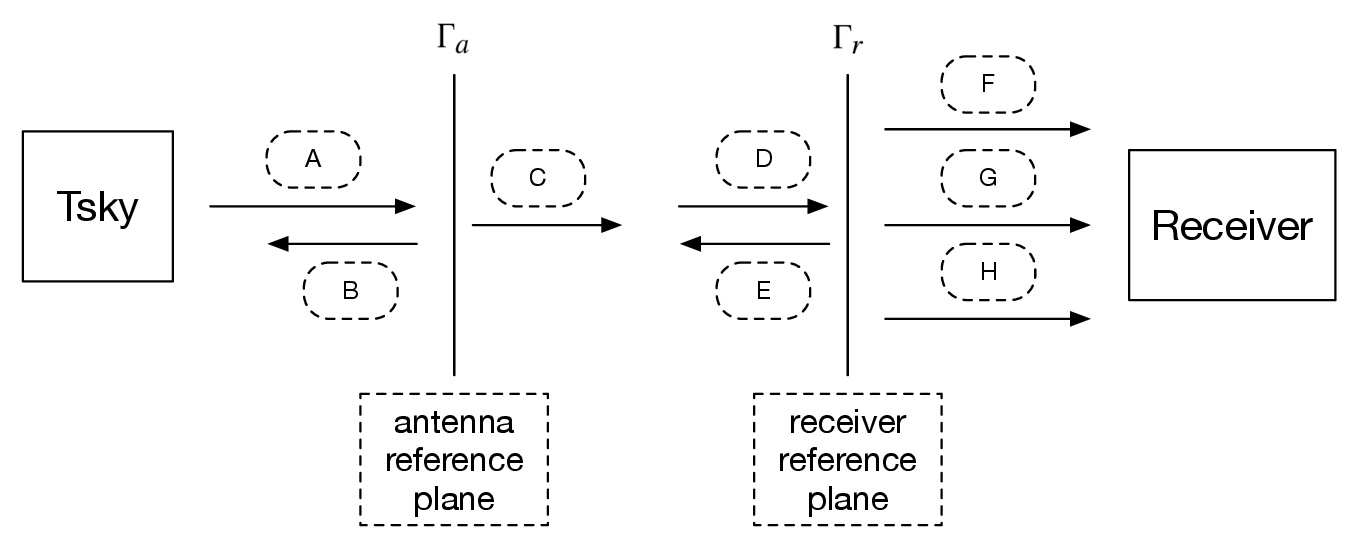}
    \caption{The model to describe transmission of noise waves between sky temperature and antenna.}
    \label{fig:appendix noise wave}
\end{figure}

The noise wave formulation was given in \citep{1129303}; here, we provide a detailed derivation in a simple model, 
as shown in Figure~\ref{fig:appendix noise wave}. We treat the sky  as an active terminator connected to an antenna. The antenna reflection coefficient $\Gamma_a$ is defined with respect to a reference plane, where all reflections of the antenna take place. Similarly, the receiver reflection coefficient $\Gamma_r$ is also defined with respect to a receiver reference plane, which can be seen as the front end of the receiver and where all the reflections occur. The whole receiver is an active device, so it does not only receive the noise passively, but also emits its own noise~wave. 

In this model, $A$ denotes the noise wave that travels forward from the sky toward the antenna reference plane, $B$ denotes the noise wave that reflects back due to antenna reflection coefficient $\Gamma_a$. $C$ denotes the noise wave that travels through the antenna reference plane; $D$ denotes the noise wave injected to the receiver reference plane that originates from $C$ and with multiple reflections at the antenna and receiver reference plane. $E$ denotes the noise wave emitted from the receiver toward the antenna. $F$ denotes the noise wave that travels through the receiver reference plane and originates from $D$; $G$ denotes another part of the noise wave that travels through the receiver reference plane due to the reflected back noise wave $E$, and $H$ denotes the self-emitted noise wave from the receiver. 

We define $U_A$, $U_B$, and $U_C$ as the noise wave voltage associated with $A, B, C$ at the antenna reference plane, and $I_A$, $I_B$, and $I_C$ as the associated currents, and $P_A$, $P_B$, and $P_C$ as the associated noise powers, then 
\begin{eqnarray}
U_{C}&=&U_{A}+U_{B}=U_{A}(1+\Gamma_a)\\
I_{C}&=&I_{A}(1-\Gamma_a)\\
P_{C}&=&\Re(U_{C}I_{C}^*)=P_{A}*\Re(1- \lvert\Gamma_a \rvert^2+\Gamma_a-\Gamma_a^*)\nonumber\\
&=& P_{A} (1-\lvert\Gamma_a \rvert^2)
\label{eq:pc vs pa}
\end{eqnarray}

The last equality is obtained by noting that $\Gamma_a-\Gamma_a^*$ is a pure imaginary number.

Considering a noise wave injected at the receiver reference plane  $D$, which originates from $C$, if $C$ reflects once at both the receiver reference plane and antenna reference plane, then its voltage is $U_{D}=U_C \Gamma_r\Gamma_a$. However, the noise wave may reflect back and forth multiple times between the antenna and the receiver. If, in each round, the reflection coefficients are the same and the attenuation along the path is small, the voltage is given by   
\begin{eqnarray}
U_{D}&=&U_C \left[1+(\Gamma_r\Gamma_a)+(\Gamma_r\Gamma_a)^2+(\Gamma_r\Gamma_a)^3+ \dots \right]\\
&=& \frac{U_C}{1-\Gamma_r\Gamma_a}
\end{eqnarray}

Considering a transmission line with characteristic impedance $Z_0$,
\begin{equation}
    \begin{aligned}
        U_{F}=\frac{U_C}{1-\Gamma_a\Gamma_r}(1+\Gamma_r), \qquad   I_{F}=\frac{1}{Z_0}\frac{U_C}{1-\Gamma_a\Gamma_r}(1-\Gamma_r); \nonumber
    \end{aligned}
\end{equation}
then
\begin{equation}
    \begin{aligned}
        P_{F}=\Re(U_{F}I_{F}^*)=\frac{1}{Z_0}\frac{\lvert U_C \rvert^2}{\lvert 1-\Gamma_a\Gamma_r \rvert^2}(1-\lvert \Gamma_r \rvert^2)
    \end{aligned}
\end{equation}

Let $P_{C}=\frac{\lvert U_C \rvert^2}{Z_O}$,  according to Equation~(\ref{eq:pc vs pa})

\begin{equation}
    \begin{aligned}
        P_{F}=\frac{P_{A}(1-\lvert\Gamma_a \rvert^2)}{\lvert 1-\Gamma_r\Gamma_a \rvert^2}(1-\lvert \Gamma_r \rvert^2)
    \end{aligned}
\end{equation}

As previously defined, $G$ is the noise wave traveling out from the receiver reference plane due to the reflected back noise wave $E$; $H$ is the self-emitted noise wave from the receiver, when $E$ reflects infinite time at both antenna and receiver reference planes; $U_{G+H}$, $I_{G+H}$, and $P_{G+H}$ can be written as

\begin{equation}
    \begin{aligned}
        U_{G+H}=\frac{U_E\Gamma_a}{1-\Gamma_a\Gamma_r}(1+\Gamma_r) + U_H
    \end{aligned}
\end{equation}

\begin{equation}
    \begin{aligned}
        I_{G+H}=\frac{1}{Z_0}\frac{U_E\Gamma_a}{1-\Gamma_a\Gamma_r}(1-\Gamma_r) + \frac{U_H}{Z_0}
    \end{aligned}
\end{equation}

\begin{equation}
    \begin{aligned}
        P_{G+H}&=\Re(U_{G+H}I_{G+H}^*)\\
        &=\frac{U_E^2\Gamma_a^2\lvert 1-\Gamma_r^2 \rvert}{Z_0\lvert 1-\Gamma_a\Gamma_r \rvert^2}+\frac{U_H^2}{Z_0}+\frac{2}{Z_0}
        \Re[\frac{\Gamma_a}{1-\Gamma_a\Gamma_r}(U_EU_H^*)]
        \label{eq:pg plus ph}
    \end{aligned}
\end{equation}

Let $T_u=\frac{U_E^2}{Z_0}$, $T_0=\frac{U_H^2}{Z_0}$, $F=\frac{ (1-\lvert \Gamma_{\rm r}\rvert^2)^\frac{1}{2} }{ 1-\Gamma_{\rm a}\Gamma_{\rm r} }$, and $\alpha$ be the phase angle of $\Gamma_aF$. Denote the last term in Equation~(\ref{eq:pg plus ph}) as $T_{\rm term}$,
then
\begin{equation}
    \begin{aligned}
    T_{\rm term}&=\frac{2}{Z_0}\Re[\frac{\Gamma_a}{1-\Gamma_a\Gamma_r}(U_EU_H^*)]\\
    &=\frac{2}{Z_0}\frac{1}{\sqrt{1-\lvert \Gamma_r \rvert^2}}\Re[\Gamma_aF(U_EU_H^*)]\\
\end{aligned}
\end{equation}

As $\Gamma_aF=\lvert\Gamma_a \rvert \lvert F \rvert e^{i\alpha}$, and $U_EU_H^*=Qe^{i\beta}$, 
we have
\begin{equation}
    \begin{aligned}
    T_{\rm term}&=\frac{2}{Z_0}\frac{\lvert\Gamma_a \rvert \lvert F \rvert Q}{\sqrt{1-\lvert \Gamma_r \rvert^2}}\Re[e^{i(\alpha+\beta})]\\
    &=\frac{2}{Z_0}\frac{\lvert \Gamma_a \rvert \lvert F \rvert Q}{\sqrt{1-\lvert \Gamma_r \rvert^2}}[\cos\alpha\cos\beta-\sin\alpha\sin\beta]
\end{aligned}
\end{equation}

Let 
\begin{equation}
    \begin{aligned}
    T_{c}=\frac{2}{Z_0}\frac{Q}{\sqrt{1-\lvert \Gamma_r \rvert^2}}\cos\beta
    \label{eq:t cos}
\end{aligned}
\end{equation}
\begin{equation}
    \begin{aligned}
    T_{s}=-\frac{2}{Z_0}\frac{Q}{\sqrt{1-\lvert \Gamma_r \rvert^2}}\sin\beta
    \label{eq:t sin}
\end{aligned}
\end{equation}

Then 
\begin{equation}
    \begin{aligned}
    P_{F+G+H}&=P_{A}(1-\lvert\Gamma_a \rvert^2)\lvert F \rvert^2+T_u\lvert\Gamma_a \rvert^2\lvert F \rvert^2+\\
    &T_c\lvert\Gamma_a \rvert\lvert F \rvert\cos\alpha+T_s\lvert\Gamma_a \rvert\lvert F \rvert\sin\alpha+T_0
\end{aligned}
\end{equation}

As previously defined, $P_{F+G+H}$ is the total noise power within the  receiver induced by the signal from the antenna; $P_A$ denotes the antenna noise power from the sky, so $T_{\rm ant}$ can be described as
\begin{equation}
    \begin{aligned}
    T_{\rm ant}&=T_{\rm sky}(1-\lvert\Gamma_a \rvert^2)\lvert F \rvert^2
    +T_u\lvert\Gamma_a \rvert^2\lvert F \rvert^2+\\
    &T_c\lvert\Gamma_a \rvert\lvert F \rvert\cos\alpha+T_s\lvert\Gamma_a \rvert\lvert F \rvert\sin\alpha+T_0.
\end{aligned}
\end{equation}

\section[\appendixname~\thesection]{Vector Network Analyzer Measurement Uncertainty} \label{appb}

As a precision instrument, the VNA should be calibrated before it is used for measurements, and several calibration techniques have been developed \citep{ballo1998applying}. Generally this is carried out by measuring standard calibrators with known impedance, and then the result is used to correct the subsequent measurements. 
The accuracy of the measurement of a device under test (DUT) depends on the accuracy and stability of the test equipment, including both the VNA and the standard calibrator, as well as the calibration method used in conjunction with the error correction model \citep{buber2019characterizing}.
Traditional full two-port calibration utilizes three kinds of impedance standards (the short, open, and load) and one transmission standard (thru) to determine the parameters with respect to the reference plane. This procedure is known as the SOLT calibration.

\begin{figure}[H]

%\begin{adjustwidth}{-\extralength}{0cm}
%\centering %% If there is a figure in wide page, please release command \centering
	\includegraphics[width=\linewidth]{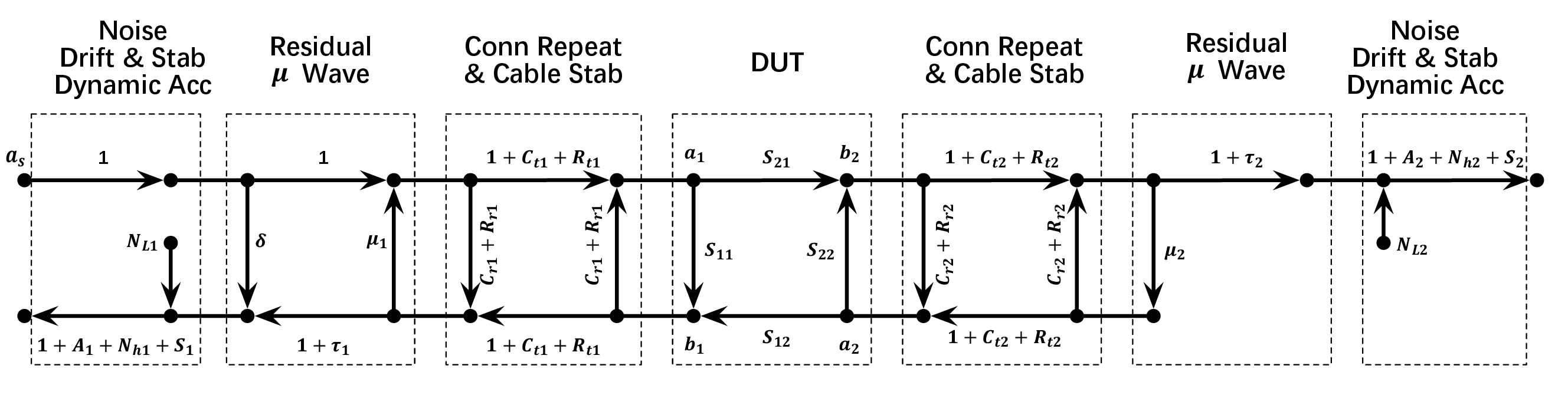}
%\end{adjustwidth}
    \caption{Signal flow graph of vector network analyzer measurement error model. A two-port device under test (DUT) is connected to the VNA, with four parameters to be measured: DUT port 1 reflection coefficient of DUT $S_{11}$, forward transmission coefficient $S_{21}$, reverse transmission coefficient $S_{12}$, port 2 reflection coefficient $S_{22}$. The VNA source port signal is denoted as $a_s$. $a_1$ and $a_2$ are the incident signals at port 1 and 2, respectively, $b_1$ is the reflected signal at port 1, and $b_2$ is the transmitted signal at port 2.  According to the characteristics of signal flow, the uncertainties can be classified into noise, drift and stability, dynamic accuracy, residual, connector repeatability, and cable stability. The residual microwave errors $\delta, \tau_1, \tau_2, \mu_1,$ and $\mu_2$ characterize the calibration standards that are not perfect. }
      \label{fig:vna signal flow graph}
      \end{figure}

The VNA measurement process is depicted in the signal flow diagram Figure~\ref{fig:vna signal flow graph}. 
In this diagram, a DUT with two ports is connected to the VNA and probed by the test signal $a_s$; the various errors are marked.
These errors can be classified into three groups: systematic, random, and drift and stability errors \citep{rytting2001network}. 
      \begin{figure}[H]
 {\captionof*{figure}{The non-linearities of the system with measurement level are described by the dynamic accuracy $A_1$ and $A_2$. These errors can be viewed as systematic errors. The low-level noise $N_{L1}$ and $N_{L2}$ of the converter determines the sensitivity of the system, and the high-level noise of the LO and IF $N_{h1}$ and $N_{h2}$ contributes to the noise on the measurement data. $C_{t1}$ and $C_{t2}$ describe the cable transmission coefficient change; $C_{r1}$ and $C_{r2}$ describe the change in the cable reflection coefficient; $R_{t1}$ and $R_{t2}$ characterize the connector transmission repeatability error; and $R_{r1}$ and $R_{r2}$ characterize the connector reflection repeatability error between calibration and measurement. These errors can be viewed as random errors. The front end and IF hardware drift with time and temperature, as characterized by the stability terms $S_1$ and $S_2$. These errors can be viewed as drift and stability. }}
\end{figure}

Systematic errors are caused by imperfections in the network analyzer and test setup. These errors are stationary and repeatable for the duration of the measurement. 
The directivity error ($\delta$) is caused primarily by coupler leakage. The tracking error ($\tau_1$ and $\tau_2$) is caused by reflectometer and mixer tracking, as well as cable length imbalance between the measurement ports. The match error ($\mu_1$ and $\mu_2$) is caused by imperfections in the calibration standards. The dynamic accuracy ($A_1$ and $A_2$) is the non-linearity of the VNA receiver over its specified dynamic range. It is a function of the power level and phase shift,  especially for high signal levels \citep{ballo1998applying}. The systematic errors can be quantified and corrected during the calibration process and mathematically reduced during measurements. However, there are always some residual systematic errors due to limitations in the calibration process, from imperfections in the calibration standards, stability and repeatability of connectors and interconnecting cables, and instrumentation drift.

Random errors vary randomly as a function of time. Noise, connector repeatability errors, and cable stability errors all contribute to random errors. Instrument noise errors include high-level noise ($N_{h1}$ and $N_{h2}$) from the local oscillator of the VNA receiver system and the low-level noise ($N_{L1}$ and $N_{L2}$) from the detector. These kind of errors have a zero mean and can be reduced by averaging multiple measurements. However, because of their random nature, noise errors cannot be mathematically corrected from a measurement. Another kind of random error is connector or switch repeatability ($R_{t1},R_{t2},R_{r1},R_{r2}$), as well as cable stability ($C_{t1},C_{t2},C_{r1},C_{r1}$). When the mechanical RF switches in the system are activated, the contacts may close differently from when they were previously activated. This can adversely affect the accuracy of the measurement. Connector repeatability errors occur because of the random variations encountered when connecting a pair of RF or microwave connectors. Variations in both reflection and transmission can be observed. Connector repeatability errors limit the achievable accuracy of all measurements. Cable stability errors are totally dependent on the quality of the test port cables used. Like connector repeatability errors, cable stability errors limit the achievable accuracy of all measurements.

Drift and stability ($S_1$ and $S_2$) errors occur when a test system’s performance changes after a calibration has been performed. The drifts are often caused by the change in environmental conditions (e.g., temperature), which affect the characteristics of the circuit components. The time frame over which a calibration remains accurate is dependent on the rate of drift that the test system undergoes in the test environment. Drift and stability errors can also be minimized by recalibration.

The total amount of these errors for $S_{11}$ measurement is given by \citep{rytting2001network}: 
\begin{equation}
\begin{aligned}
    \Delta S_{11}(\text{magnitude}) = 
    \text{Systematic} + \sqrt{\text{Random}^2 + (\text{Drift \& Stability})^2}
\end{aligned}
\label{eq:error magnitude}
\end{equation}

Equation~(\ref{eq:error magnitude}) gives the uncertainty in the magnitude of reflection coefficients, where 
the error terms are defined by their absolute magnitude. The systematic errors are added directly, while the random, 
drift and stability errors can usually be added in quadrature. 
The systematic errors can be represented as
\begin{equation}
\begin{aligned}
    \text{Systematic}=\delta + \tau_1 S_{11} + \mu_1 S_{11}^2 + \mu_2 S_{21} S_{12} + A_1 S_{11}
\end{aligned}
\label{eq:systematic}
\end{equation}
where $\delta, \tau_1, \tau_2, \mu_1,$ and $\mu_2$ characterize the imperfection of the calibration standards.
The random error can be represented as: 
\begin{equation}
    \text{Random} = \sqrt{(C_r)^2+(R_r)^2+(N_r)^2}
\end{equation}
in which
\begin{equation}
    C_r = \sqrt{(C_{r1})^2+(2C_{t1}S_{11})^2+(C_{r1}S_{11}^2)^2+(C_{r2}S_{21}S_{12})^2}
\end{equation}
\begin{equation}
R_r=\sqrt{(R_{r1}+2R_{t1}S_{11}+R_{r1}S_{11}^2)^2+(R_{r2}S_{21}S_{12})^2}
\end{equation}
\begin{equation}
    N_r=\sqrt{(N_{h1}S_{11})^2+(N_{L1})^2}
\end{equation}
where $C_{r1}$ and $C_{r2}$ describe the change in the cable reflection coefficient; $C_{t1}$ and $C_{t2}$ describe the cable transmission coefficient change; $R_{t1}$ and $R_{t2}$ 
characterize the connector transmission repeatability error; $R_{r1}$ and $R_{r2}$ characterize the connector reflection repeatability error between calibration and measurement. $N_{L1}, N_{L2}, N_{h1},$ and $N_{h2}$ characterize 
the noise in the VNA system. 
The drift and stability can be represented as 
\begin{equation}
    \text{Drift \& Stab} =S_1S_{11}
\end{equation}
where $S_1$ characterizes the front end drift with time and temperature.
The reflection coefficient phase uncertainty can be represented as
\begin{equation}
    \text{phase}(\Delta S_{11})=\arcsin(\frac{\Delta S_{11}(\text{magnitude})}{S_{11}(\text{magnitude})})+2C_{t1}+S_1.
    \label{eq:error phase}
\end{equation}

The parameters of the VNA error model are provided by the manufactures.

%%%%%%%%%%%%%%%%%%%% bibliography %%%%%%%%%%%%%%%%%%%%%
%%%%%%%%%%%%%%%%%%%% bibliography %%%%%%%%%%%%%%%%%%%%%
%%%%%%%%%%%%%%%%%%%% bibliography %%%%%%%%%%%%%%%%%%%%%
%%%%%%%%%%%%%%%%%%%% bibliography %%%%%%%%%%%%%%%%%%%%%
%%%%%%%%%%%%%%%%%%%% bibliography %%%%%%%%%%%%%%%%%%%%%
%%%%%%%%%%%%%%%%%%%% bibliography %%%%%%%%%%%%%%%%%%%%%
%%%%%%%%%%%%%%%%%%%% bibliography %%%%%%%%%%%%%%%%%%%%%
%%%%%%%%%%%%%%%%%%%%%%%%%%%%%%%%%%%%%%%%%%
\begin{adjustwidth}{-\extralength}{0cm}
%\printendnotes[custom] % Un-comment to print a list of endnotes

\printendnotes[custom]

\reftitle{References}

\PublishersNote{}
\end{adjustwidth}

\end{document}